\documentclass[acmsmall,screen,nonacm]{acmart}

\setcopyright{none}
\makeatletter
\@ifundefined{hyxmp@parse@acmart}{}{\let\hyxmp@parse@acmart\relax}
\makeatother

\usepackage{amsmath}
\usepackage{mathtools}
\usepackage{booktabs}
\usepackage{tabularx}
\usepackage{array}
\usepackage{multirow}
\usepackage{enumitem}
\usepackage{graphicx}
\usepackage{xcolor}
\usepackage{tikz}
\usepackage{url}
\usepackage{microtype}
\usepackage{algorithm}
\usepackage{algpseudocode}
\usepackage{placeins}
\usetikzlibrary{arrows.meta,positioning,fit,calc,shapes.geometric,backgrounds}

\hypersetup{
  hidelinks,
  keeppdfinfo=true,
  pdftitle={Authorization for Self-Modifying AI Agent Populations: Conserving Authority across Replacement, Forking, and Rollback},
  pdfauthor={Genliang Zhu; Chu Wang},
  pdfsubject={Population-safe authorization succession across self-modifying agent generations},
  pdfcreator={LaTeX}
}

\setlist[itemize]{leftmargin=1.25em}
\setlist[enumerate]{leftmargin=1.45em}
\newcolumntype{L}[1]{>{\raggedright\arraybackslash}p{#1}}
\newcolumntype{Y}{>{\raggedright\arraybackslash}X}

\definecolor{lfBlue}{HTML}{1D4ED8}
\definecolor{lfBlueFill}{HTML}{EFF6FF}
\definecolor{lfGreen}{HTML}{15803D}
\definecolor{lfGreenFill}{HTML}{F0FDF4}
\definecolor{lfAmber}{HTML}{B45309}
\definecolor{lfAmberFill}{HTML}{FFFBEB}
\definecolor{lfPurple}{HTML}{6D28D9}
\definecolor{lfPurpleFill}{HTML}{F5F3FF}
\definecolor{lfRed}{HTML}{B91C1C}
\definecolor{lfRedFill}{HTML}{FEF2F2}
\definecolor{lfSlate}{HTML}{475569}

\newtheorem{definition}{Definition}
\newtheorem{lemma}{Lemma}
\newtheorem{theorem}{Theorem}

\newtheorem{corollary}{Corollary}

\newcommand{\mcode}[1]{\ifmmode\text{\normalfont\ttfamily #1}\else{\normalfont\ttfamily #1}\fi}

\newcommand{\Active}{\ensuremath{\mathsf{Active}}}
\newcommand{\Fenced}{\ensuremath{\mathsf{Fenced}}}
\newcommand{\Quarantined}{\ensuremath{\mathsf{Quarantined}}}

\newcommand{\Allow}{\textsc{Allow}}
\newcommand{\Deny}{\textsc{Deny}}
\newcommand{\Indeterminate}{\textsc{Indeterminate}}

\title[Authorization for Self-Modifying AI Agent Populations]
{Authorization for Self-Modifying AI Agent Populations:\\
Conserving Authority across Replacement, Forking, and Rollback}

\author{Genliang Zhu}
\affiliation{%
  \institution{Accentrust}
  \city{Vancouver}
  \country{Canada}
}
\affiliation{%
  \institution{Georgia Institute of Technology}
  \city{Atlanta}
  \state{Georgia}
  \country{USA}
}
\email{research@accentrust.com}

\author{Chu Wang}
\affiliation{%
  \institution{Accentrust}
  \city{Vancouver}
  \country{Canada}
}
\affiliation{%
  \institution{University of Illinois Urbana-Champaign}
  \city{Urbana}
  \state{Illinois}
  \country{USA}
}

\begin{abstract}
Self-modifying agents can replace, fork, and roll back identity-bearing
software while descendants remain executable. Per-successor authorization
does not constrain the population: siblings can duplicate quotas, combine
complementary permissions, survive ancestor cuts, or overlap predecessors
during promotion. We define authorization succession across software
generations, which conserves authority across a single-parent generation
forest's active frontier.

We present an external protocol that binds each generation to a
manifest, root, unique parent, complete lineage, and fresh population sequence.
Separate invariants jointly bound root-lifetime consumption and current
population exposure.
A staged reservation freezes the predecessor residual for replacement; a
partitioning fork validates its full child family inline. Each commit atomically
fences the predecessor and activates its successors. Ancestor cuts invalidate
dependent descendants; rollback creates a fresh generation without restoring
spent authority; a new root requires an independent grant. With complete mediation,
authenticated records, sound effect abstraction, durable monotone state, and
complete lineage accounting, we prove population-safe succession, fork
conservation, revocation closure, atomic handoff, rollback non-reminting, and
exclusion of self-certification.

An executable evaluation matches 32 registered decisions through
direct-call and mailbox mappings (64/64 replays; 28 allows, 36 denies). An
independent checker accepts all 64 original traces and rejects 28/28
outer-resealed semantic mutants; 12/12 profile invariants, 16/16 crash cuts,
and 32/32 contender schedules pass. Separate adapters reproduce all 32
decisions around one measured OurArk descendant and one Darwin G\"odel Machine
EditTool mutation, each with fresh-process restart, atomic succession, and
predecessor rejection. These results establish
authorization succession over registered protected effects, not arbitrary
program equivalence or unmediated behavior.
\end{abstract}


\begin{document}
\maketitle
\hypersetup{
  keeppdfinfo=true,
  pdftitle={Authorization for Self-Modifying AI Agent Populations: Conserving Authority across Replacement, Forking, and Rollback},
  pdfauthor={Genliang Zhu; Chu Wang},
  pdfsubject={Population-safe authorization succession across self-modifying agent generations},
  pdfcreator={LaTeX}
}

\section{Introduction}
\label{sec:introduction}

Self-modifying agents make authorization a population problem. A running agent
can edit its code, change its model route or tool harness, create an isolated
candidate, promote a successor, retain several successful branches, merge
their changes, and later roll back to an earlier artifact. Each resulting
process may present a valid-looking identity and may individually satisfy a
transition policy. Yet authorization is exercised by the complete set of
simultaneously live generations. Treating each accepted successor as a fresh
copy of the predecessor's grant silently multiplies budgets, concurrency
slots, approval weight, and effect combinations.

Existing controls establish important adjacent facts. Versioned agent bodies
make descent and inherited validation inspectable \cite{ourark2026}.
Self-improving systems demonstrate open-ended code mutation and selection
across an archive of agents \cite{dgm2025}. Mutation-aware authorization can
decide whether one changed agent remains inside a grant-fixed transition
envelope \cite{earnedauthority2026}, while capability-bound certificates can
invalidate a credential when its model/tool/permission manifest changes
\cite{dynamiccapabilities2026}. Delegation systems can attenuate
capabilities and revoke workflow edges \cite{boundedagents2026,vera2026}, and
recursive-tree controls can conserve probabilistic risk or deterministic
resources \cite{prv2026,agentcontracts2026,topologybudget2026,tokenbudgets2026}.
An external safety kernel can force protected effects through an immutable
decision path \cite{unfireable2026}. None of these facts alone specifies how an
authenticated software-generation lifetime is atomically replaced, how its
population allocation is joined to a lineage cut and independent re-root, or
whether a returned historical body has replayed that generation's retired
grant. The remaining object addressed here is this authorization-bearing
generation frontier and its succession transaction.

\subsection{A strict population counterexample}

Suppose generation $g_0$ holds authority to submit payments satisfying a
typed relation $E_0$, spend at most 100 units, and occupy one production
executor slot. It produces two candidates, $g_1$ and $g_2$. Both make benign
code changes and independently remain inside the same transition envelope.
A per-successor rule therefore assigns each branch the grant of $g_0$.
The pair can now spend 200 units and execute concurrently in two slots. If one
branch retains a sensitive-data read projection and the other retains an
external-destination projection, later coordination or code merging can also
synthesize an effect tuple that no authority witness authorized. Every local transition was acceptable;
the population transition was not.

Revocation and rollback expose the same integration gap. Hierarchical
credential and delegation systems already provide descendant or edge
revocation \cite{hbhc2026,vera2026}, and checkpoint controls address authority
resurrection \cite{acrfence2026,agentlibos2026}. Revoking $g_0$ while $g_2$ is
offline does not necessarily revoke a cached grant later presented by a
grandchild. Replacing $g_0$ with $g_1$ in two separately committed steps leaves
a crash window in which both can submit. Checking out the byte-identical body
of $g_0$ can resurrect its historical identifier unless the generation
sequence and authorization lineage are distinct from body content. These executions are
not failures of mutation classification. They are failures to conserve and
retire authority across a changing generation graph.

\subsection{Authorization succession across generations}

This paper defines authorization succession across software generations. A
generation is an authenticated principal whose identity binds its body,
security-relevant runtime configuration, unique authorization parent, and
monotonically fresh generation sequence. Each authorization root has a lifetime
grant containing typed effect relations, consumable budgets, and concurrency
shares; the population separately retains a persistent ceiling on current
exposure. The registered succession profile also fixes evidence-independence
requirements. Descendants receive ledger allocations within both bounds; a fork
reallocates existing authority instead of copying it.

We present an external protocol for maintaining these objects.
Candidate generations remain quarantined while independent evidence is
collected. The ledger reserves an exact residual authority slice and commits
promotion at one linearization point. A replacement commit simultaneously
fences one predecessor and activates one successor; a partitioning fork fences
one predecessor while atomically activating two or more single-parent children
whose allocations fit its residual. Every
protected effect is checked against the exact current generation, current
authorization root, allocation, population residual, and lineage cut set.

Ancestor revocation creates a lineage cut. Every descendant whose authority
witness still contains the revoked atom becomes ineligible, including delayed
and nested descendants that start later. A descendant may continue only by an
explicit independent re-root transaction supported by an authority root that
does not derive from the revoked lineage. Rollback creates a new generation
with a fresh population-issued sequence and new authorization record even when its body digest equals a
historical digest. Body equality can justify code reuse; it cannot remint old
authority.

\subsection{Contributions}

This paper makes five contributions:

\begin{enumerate}
  \item It defines authorization succession over an authenticated software-
  generation population and gives strict separating executions for sibling duplication,
  relational permission splicing, ancestor-revocation leakage, dual-active
  promotion, rollback replay, and self-certification. These executions remain
  possible under a per-generation transition ceiling.

  \item It introduces an authorization model over an authenticated generation
  lineage forest. A root grant bounds lifetime authority, including irreversible
  debit, while a persistent population ceiling independently bounds current
  relational effect authority, additive budgets, and live-executor shares.
  Promotion separately requires evidence-root independence.

  \item It presents a generation-aware protocol with quarantined candidates,
  independent evidence records, staged exact reservations, inline family
  authorization for forks, atomic commits, durable predecessor fences, complete ancestor cuts, explicit
  independent re-root, and rollback-as-new-generation semantics.

  \item It proves conditional population-safe succession, fork conservation,
  cross-generation revocation closure, atomic generation handoff, rollback
  non-reminting, and exclusion of self-certification. The premises are stated
  as enforceable registry, ledger, evidence, and effect-mediation obligations.

  \item It supplies a registered executable evaluation over mutation and
  lineage cases, two separately implemented logical runtime mappings, durable
  crash cuts, deterministic interleavings, an independent trace checker, and
  consistently resealed semantic regressions. Two adapters to frozen public
  runtimes separately exercise source-native descent or mutation, fresh-process
  restart, succession, predecessor rejection, and the complete registered
  replay matrix. Each empirical statement is tied to the exact finite evidence
  denominator that produced it.
\end{enumerate}

\subsection{Assurance scope}

The protected property is operation-time authorization at registered effect
brokers. It covers generations, evidence and authorization records, lineage cuts,
allocations, and effects whose security-relevant fields cross those brokers.
It establishes authority conservation and revocation for the declared
generation graph, root grants, and persistent population ceiling. Semantic equivalence of arbitrary
programs, correctness of an evaluator's substantive judgment, effects outside
the registered mediation boundary, and reversal of already committed physical
outcomes are distinct properties with distinct evidence requirements.

The remainder of the paper separates population succession from adjacent
controls (Section~\ref{sec:background}), states the system and assurance model
(Section~\ref{sec:system}), formalizes the protocol and proves its properties
(Section~\ref{sec:formal}), presents the design and logical instantiations
(Sections~\ref{sec:design}--\ref{sec:instantiations}), evaluates the registered
artifact and refinements over frozen public runtimes (Sections~\ref{sec:evaluation}--
\ref{sec:evidence}), and closes with related work and the exact assurance
boundary.

\section{Problem Setting and Separating Executions}
\label{sec:background}

Authorization succession sits at the intersection of versioned identity,
capability transfer, revocation, and distributed cutover. This section fixes
the distinctions on which the protocol depends.

\subsection{Artifact identity is not generation authority}

A body digest can identify the files, prompts, policies, skills, and harness
used by an agent. It does not uniquely identify one authorization lifetime.
The same body can be installed twice, restored after revocation, or launched
under two different roots. Conversely, a body may undergo a semantics-
preserving rebuild while remaining the intended successor. The proposed model
therefore separates:

\begin{itemize}
  \item a \emph{body commitment}, which binds security-relevant executable
  content and configuration;
  \item a \emph{generation principal}, which binds that commitment to one
  authorization parent (or no parent for an independently issued root), a
  lineage root, a fresh generation sequence, and its creation transaction; and
  \item an \emph{authority allocation}, which names the exact effect relation,
  budgets, shares, and evidence constraints currently assigned to that
  generation.
\end{itemize}

Content equality may support a transition decision, but it never aliases two
generation principals. This distinction is what makes replayed rollback and
concurrent identical workers observable to the authorization layer.

\subsection{Lifecycle acceptance is not population conservation}

A mutation evaluator answers whether a candidate satisfies a transition
contract. A promotion controller answers which candidate becomes live. A
protected-effect broker answers whether a particular operation may cross its
commitment point. Population safety connects all three decisions through a
single durable allocation state.

Branch-level conservation is already established for probabilistic risk,
multidimensional resources, dynamic multi-agent topologies, and affine token
budgets \cite{prv2026,agentcontracts2026,topologybudget2026,tokenbudgets2026}.
The succession setting adds an authorization-bearing software generation,
predecessor retirement, lineage cuts, and rollback. Checking candidates one at
a time loses information. If two successors each
pass the same ceiling check, a Boolean capability copied to both may increase
parallel exercise, an additive budget may double, and two partial relational
permissions may later be joined. The authorization algebra must therefore
preserve multiplicity and tuple correlations. We model a protected effect as a
complete typed tuple---for example, operation, resource, destination, purpose,
provenance, amount class, and risk class---rather than as independently
projected fields. Authenticated authorization objects allocate complete relations and
explicit quantitative shares.

\subsection{Succession operations}

The generation graph supports four operations with different authorization
semantics:

\begin{description}[leftmargin=6.8em,style=nextline]
  \item[Replacement.] One active predecessor is fenced and one successor is
  activated in a single durable commit. No protected-effect check can observe
  both the old and new allocation as active.

  \item[Partitioning fork.] One active predecessor is fenced while two or more
  single-parent children are activated atomically. Their additive allocations
  are bounded by the predecessor's unspent residual, and a registered
  single-holder effect atom may be assigned to at most one child. Forking
  creates principals, not authority.

  \item[Branch coordination.] Siblings may later coordinate or feed a common
  downstream action, but such coordination does not create an authorization
  parent edge or union their grants. The complete downstream effect must be
  admitted against one current authority witness or against a newly issued
  independent root. Multi-parent authority merge is outside the core
  transition system studied here.

  \item[Rollback.] A historical body commitment may seed a new successor, but
  the successor receives a fresh population-issued generation sequence and a
  new authorization decision. Historical evidence and authorization records,
  reservations, receipts, and spent budget stay
  terminal.
\end{description}

An \emph{independent re-root} is not an additional graph-editing shorthand. It
is an explicit authority transaction that supplies a fresh external root,
proves that the new grant does not derive from the old lineage, allocates a new
envelope, and removes the old lineage atom from the continuing generation's
authority witness.

\subsection{Six strict separating executions}

Table~\ref{tab:separating-executions} states the minimum executions that a
population-safe mechanism must distinguish. Each row contains locally valid
artifacts; the violation appears only when lineage and population state are
considered together.

\begin{table*}[t]
\caption{Strict executions separating population-aware succession from per-generation acceptance.}
\label{tab:separating-executions}
\footnotesize
\begin{tabularx}{\textwidth}{L{0.18\textwidth}L{0.26\textwidth}Y L{0.20\textwidth}}
\toprule
Execution & Locally acceptable observation & Population violation & Required state \\
\midrule
Sibling duplication & Both children remain below the parent's individual ceiling & The pair receives two copies of one budget or executor share & Aggregate allocation and residual \\
Relational splice & Each branch exposes only an allowed field projection & Branch coordination synthesizes a complete effect tuple never certified by one authority witness & Complete effect relation and holder ownership \\
Ancestor-revoke leak & Offline descendant presents a previously valid child authorization record & Its witness still depends on the revoked ancestor atom & Lineage cut and complete frontier \\
Dual-active promotion & New generation activates; predecessor fence is retried after a crash & Old and new generations both reach the protected sink & Atomic replacement commit \\
Rollback replay & Historical body digest and old authorization record match & Retired sequence, spent budget, or revoked grant is reused & Fresh generation sequence and monotone identity/state guards \\
Self-certification & Distinct certifier identifiers approve a candidate & All certifiers are declared under the candidate's control root & Evidence-root independence \\
\bottomrule
\end{tabularx}
\end{table*}

\subsection{Design requirements}

The separating executions yield seven requirements. Generation identity must
be exact and fresh. Every active descendant must be discoverable from a
durable lineage frontier. Authority allocation must be aggregate and
relation-preserving. Promotion must bind evidence that is independent of the
candidate's control root. Replacement and allocation changes must be atomic at
one ledger linearization point. Ancestor revocation must reach delayed and
offline descendants at their next mediated transition. Finally, every
protected effect must validate current lineage and allocation state at the
commit boundary; a cooperative process promise is insufficient.

These requirements explain the paper's boundary with adjacent systems.
Mutation classification supplies evidence to succession, versioned bodies
supply identity inputs, delegation supplies attenuation rules, and an external
kernel supplies a trustworthy path. The population ledger adds the state that
none of those components can reconstruct after authority has been copied into
independent sessions.

\section{System, Threat, and Assurance Model}
\label{sec:system}

\subsection{Assurance statement}

We consider a self-modifying agent population derived from an authorization
root $r$. The protected statement is that every effect accepted at time $t$ is
attributed to exactly one current active generation with a complete authority
witness, that the aggregate authority allocated to all active dependent
descendants remains below both $r$'s lifetime grant and the population's
persistent current-exposure ceiling, and that no descendant
can exercise an ancestor atom after its durable revocation cut. Replacement
promotion additionally establishes that predecessor fencing and successor
activation share one linearization point. A body restored by rollback is a
fresh generation and receives no authority from historical identity alone.

The assertion is evaluated at the protected-effect commitment point. Planning,
candidate generation, evaluation, and code execution may be adversarial; they
cannot turn an uncommitted proposal into a protected effect without a current
ledger admission. The statement remains true during crashes and partitions by
denying or retaining an indeterminate/quarantined state until authenticated
current evidence is available.

\subsection{Components and trusted boundary}

The system has six logical components:

\begin{enumerate}
  \item The \emph{generation registry} stores content-addressed generation
  identities, each generation's unique authorization parent, lineage roots,
  epochs, phases, and immutable
  creation records.

  \item The \emph{population ledger} owns root grants, persistent population
  ceilings, active-generation
  allocations, reservations, residual budgets, revocation cuts, and replay
  tombstones. It is the linearization authority for every population change.

  \item A \emph{candidate runtime} creates and evaluates modifications in
  quarantine. Its own report may be evidence input but cannot activate itself.

  \item Independent \emph{evidence roots} attest registered predicates about
  body identity, effect-envelope compatibility, validation results, and
  operator approval. Independence is evaluated by authenticated control root,
  not by display name or process identifier.

  \item The \emph{succession coordinator} proposes reservations and commits,
  but has no authority to bypass ledger predicates. Only protected-effect
  requests promise exact-result retry under A10. Other repeated commands are
  accepted or rejected by their phase, uniqueness, version, and monotonicity
  guards; they do not inherit an exact-retry guarantee.
  
  \item A \emph{protected-effect broker} validates generation identity,
  complete lineage atoms and the current cut set, active phase, allocation,
  complete typed effect, budget, and
  revocation frontier in the same serialized operation that reserves or
  commits the effect.
\end{enumerate}

The registry, population ledger, admitted evidence/control-root bindings, and effect broker lie
outside the mutable generation boundary. They can be implemented by separate
processes or one isolated service; the logical trust separation and durable
state obligations are invariant. An out-of-process only-path kernel is a
compatible substrate for these components
\cite{saltzer1975,unfireable2026,runtimegovernance2026,crosssubstrate2026}.

\subsection{Adversary and fault model}

The adversary controls every candidate and active generation, including its
planner, model route, code, Skill closure, memory policy, policy compiler, and
runtime configuration. It may fork repeatedly, construct deep lineages,
submit an unsupported multi-parent merge, retain old credentials, delay an
offline descendant, alias certifier principals under one control root, present
partial effect projections, reorder promotion messages,
replay certificates and receipts, race predecessor and successor effects, and
crash the coordinator at every durable boundary. It may restore any historical
body and may supply a byte-identical artifact under a new launch.

The adversary cannot cause an authenticated root-authorization or certification
predicate to accept without an admitted external control root, find a collision
in the registered canonical digest, rewrite a committed ledger record, bypass
every registered protected-effect broker, or violate the atomicity and
serializable ordering of the trusted ledger. A compromised broker, omitted
effect path, falsely admitted evidence root, incomplete lineage registration,
or unsound effect normalizer violates an explicit premise; such evidence
cannot be converted into a theorem conclusion. Authentication is an abstract
premise. The reference prototype validates a frozen declared-principal-to-
control-root mapping and digest-bound records; it does not authenticate the
caller, prove key possession, or implement public-key signatures. Those
authentication obligations remain part of A2 for a deployment satisfying the
theorem.

Networks and candidate runtimes may be unavailable for an unbounded interval.
Safety requires no delivery deadline. Eventual activation additionally
requires eventual delivery, stable current evidence, and available durable
storage; it is outside the safety result stated here.

\subsection{Assurance assumptions}

Table~\ref{tab:lf-assumptions} gives the premises used by the formal results.
A1--A10 are safety assumptions.

\begin{table*}[t]
\caption{Assurance assumptions for a registered succession profile.}
\label{tab:lf-assumptions}
\footnotesize
\begin{tabularx}{\textwidth}{L{0.05\textwidth}L{0.205\textwidth}Y}
\toprule
ID & Assumption & Enforced obligation \\
\midrule
A1 & Complete effect mediation & Every security-relevant commitment named by the profile crosses a broker that validates the current generation, complete lineage-atom closure, allocation, and complete effect tuple at its own linearization point. \\
A2 & Authentic canonical identity & Generation identities, the unique authorization parent, body/configuration commitments, authorization and certification records, cuts, allocations, receipts, and evidence-root identities use unambiguous canonical encodings and authenticators whose acceptance cannot be produced from the candidate's control root alone. \\
A3 & Durable monotone generations & Generation and root epochs never decrease or reactivate; accepted evidence/authorization records, lineage cuts, fences, root-lifetime debit records, and effect-request replay records cannot be rewritten. \\
A4 & Complete lineage accounting & Every non-root generation has exactly one registered authorization parent before activation. A fork may create several successors atomically, but each successor names the same single predecessor; the predecessor is fenced. The ledger retains every resulting parent edge and lineage atom across offline, delayed, and nested descendants until a terminal cut or independent re-root removes authority. \\
A5 & Sound population algebra & Effect normalization preserves every policy-relevant correlation; the registered effect-membership predicate and authority join are monotone, multiplicity-aware where policy requires it, additive for consumable resources, and conservative for unknown fields. \\
A6 & Atomic population transactions & Reservation, allocation replacement, activation, predecessor fencing, and budget consumption execute as serializable durable transitions with exact population-version and source-residual preconditions. A held reservation either blocks predecessor consumption or is made stale by it. \\
A7 & Evidence-root independence & Each promotion profile names required predicates and independent control-root classes; aliases, descendants, cloned processes, and principals governed by the candidate's control root cannot satisfy independent slots. \\
A8 & Transfer non-amplification & An accepted authorization record binds one exact predecessor, one successor or atomic successor set, the predecessor's reconstructed residual, successor grants, population state, manifest commitments, evidence records, and consumed identifiers. The ledger accepts only when the exact post-state remains below the predecessor residual, root-lifetime grant, and persistent population ceiling. \\
A9 & Complete revocation fencing & An ancestor cut is checked on activation, succession, fork, re-root, rollback, and every protected effect. Cached child records cannot override a monotone lineage cut. \\
A10 & Effect-request idempotency & For protected-effect requests only, an exact retry under the same request identity returns the previously committed effect receipt without another debit; the same identity with a different canonical effect request is rejected. No exact-retry guarantee is assumed for proposal, certification, reservation, succession, fork, cut, re-root, or administrative commands; their replay safety follows from A3 and A6 phase, uniqueness, and monotonicity guards. \\
\bottomrule
\end{tabularx}
\end{table*}

\subsection{Registered profile and decision surface}

A profile fixes the generation-identity schema, effect normalizer, authority
algebra, evidence-root independence classes, permitted succession operations,
timeout semantics, and all brokers reachable by the root. A general deployment
may bind these choices with a separate profile digest. The executable reference
profile instead fixes its schema in the implementation and binds six manifest
digests plus the external-control path and proposer identity; it implements
single-parent succession, single-predecessor fork, rollback, cuts, and
independent re-root, but no multi-parent merge. Unknown identity fields,
incomplete lineage, missing evidence, unavailable current allocation state,
and conflicting records yield \Indeterminate{} or remain \Quarantined; they
never become \Allow{} through timeout.

The broker's decision surface is deliberately small. It receives a generation
principal, a complete normalized effect, declared costs, and a retry-stable
effect-request identity. Within its serialized ledger operation it loads the
authoritative population and generation records, then checks the active state,
lineage cut set, allocation, held reservations, and remaining budget.
Current-root consistency follows from the ledger's population-transition
invariant rather than a caller-supplied version. It returns an accepted effect
receipt, the same receipt for an exact effect retry, or a typed rejection. The
agent cannot choose which identity fields, authorization parent, budgets, or
effect dimensions participate in the decision.

\subsection{Assurance boundary}

The theorems quantify over the registered generation graph, root grants,
population ceiling, normalizer, evidence profile, and mediated effects. They establish enforcement
of the supplied policy relation and lineage state; they do not infer whether
arbitrary code is semantically equivalent or whether an evaluator's external
factual judgment is correct. Evidence admission and profile governance supply
those inputs. The design keeps this boundary explicit so a strong ledger
result cannot be mistaken for a code-understanding or deployment-universality
claim.

\section{Formal Model}
\label{sec:formal}

This section defines authorization succession over a population of mutable
agent generations. The model separates content identity, authorization
lineage, and artifact provenance: reusing bytes does not reuse authority, and
copying a predecessor does not copy its allocation. The six stated results are
proved in Appendix~\ref{app:proofs}.

\subsection{Assumptions, abstraction, and trusted boundary}

The authoritative premises are A1--A10 in
Table~\ref{tab:lf-assumptions}: complete effect mediation, authentic canonical
identity, durable monotone generations, complete lineage accounting, sound
population algebra, atomic population transactions, evidence-root
independence, transfer non-amplification, complete revocation fencing, and
effect-request idempotency. The trusted-boundary definition in
Section~\ref{sec:system} additionally places the lineage authority, frozen
control-root registry, authenticated-record validators, and durable ledger
outside every candidate generation's mutable closure. A later delivery or
physical consequence of a pre-cut commitment remains attributed to its unique
pre-cut admission event.

The theory is parameterized by a registered identity profile, authority
algebra, and authenticated-record predicates. Those parameters support a
general assurance statement; they are not claims that the executable artifact
implements every possible profile. The reference realization fixes the
manifest schema, uses effect-label sets and nonnegative integer budgets, and
validates declared control-root separation and digest-bound authorization
records. Caller authentication remains an A2 deployment premise; the artifact
does not implement public-key signature verification,
multi-parent generation identity, or a merge transition. The model likewise
does not infer arbitrary program equivalence, discover unmanifested effect
paths, or establish the truth of a certifier's observations merely because an
authenticated record was accepted.

\subsection{Generation identity and authorization lineage}

Let \(\mathcal P\) be the set of governed populations, \(\mathcal R\) the set
of authorization roots, \(\mathcal G\) the set of generations, and
\(\mathcal E\) the set of normalized protected effects. Each root
\(r\in\mathcal R\) is bound to one population \(p(r)\), one monotone root
epoch \(\epsilon_r\), and an authenticated root-grant digest \(h^G_r\). Every
non-root generation has one unique authorization parent
\(\mathsf{par}(g)\in\mathcal G\); a root generation has
\(\mathsf{par}(g)=\bot\). Each generation \(g\in\mathcal G\) is registered
under exactly one authorization root, written
\(\mathsf{root}(g)\in\mathcal R\). If \(\mathsf{par}(g)\ne\bot\), then
\(\mathsf{root}(\mathsf{par}(g))=\mathsf{root}(g)\), so every ordinary parent
edge remains within one root lifetime. Each population owns a monotone
generation counter and records exactly one current
authorization root \(\mathsf{curRoot}_S(p)\), with
\(p(\mathsf{curRoot}_S(p))=p\). A re-root may advance that pointer only through
Definition~\ref{def:reroot}; historical root records remain immutable.

Let \(M_g\) denote the canonical registered manifest and
\(m_g=H(\mathsf{Enc}(d_M,M_g))\). The executable reference manifest consists
exactly of the named fields
\[
 \begin{split}
 M_g^{\rm ref}=\{&\mathsf{codeDigest},\mathsf{controlPath},
 \mathsf{enforcementConfigDigest},\mathsf{memoryPolicyDigest},\\
 &\mathsf{modelRouteDigest},\mathsf{policyCompilerDigest},
 \mathsf{proposerId},\mathsf{skillClosureDigest}\}.
 \end{split}
\]
All eight values are nonempty and unknown or missing fields are rejected. A
general registered profile may define a larger manifest and bind its profile
identifier within that manifest. Results for such a profile are conditional
on A2 and its registered resolver; the reference-artifact evidence covers only
the fixed schema above.

Let \(a_r\) be the lineage atom authenticated by the root grant. For a
generation under root \(r\), define its self atom, complete lineage-atom set,
and generation identifier by
\begin{equation}
\label{eq:generation-identity}
\begin{aligned}
 a_g={}&H(\mathsf{Enc}(d_{\rm atom},p(r),r,\epsilon_r,h^G_r,n_g,
                 \mathsf{par}(g),m_g)),\\
 A_g={}&
 \begin{cases}
   \{a_r,a_g\}, & \mathsf{par}(g)=\bot,\\
   A_{\mathsf{par}(g)}\cup\{a_g\}, & \text{otherwise},
 \end{cases}\\
 \mathsf{gid}(g)={}&H(\mathsf{Enc}(d_{\rm gen},p(r),r,\epsilon_r,h^G_r,n_g,
                 \mathsf{par}(g),\mathsf{sort}(A_g),m_g)).
\end{aligned}
\end{equation}
Here \(n_g\) is a fresh population-issued sequence number. Thus equal content
does not imply equal generation identity, and a fresh re-root cannot collide
with an old authorization lifetime. A separately recorded
\(\mathsf{source}(g)\) may name copied code, checkpoints, or artifacts; it is
provenance only and never an authorization parent. A fork atomically creates
siblings \(N\), each satisfying \(\mathsf{par}(g)=p\) for the same unique
predecessor \(p\). The core transition system admits no multi-parent child and
no merge rule.

Write \(p\rightarrow_L g\) when \(p=\mathsf{par}(g)\), and let
\(\mathsf{Anc}_L(g)\) be the transitive closure of this relation. For each
population \(p\), the ledger stores a monotone, population-wide set
\(\mathsf{Cut}_S(p)\) of cut lineage atoms. The set is retained across an
independent re-root, so changing the current root cannot erase a historical
revocation. For \(r=\mathsf{root}(g)\), define
\begin{equation}
\label{eq:current}
 \begin{split}
 \mathsf{Current}_S(g)\iff{}&
   r=\mathsf{curRoot}_S(p(r))\ \wedge\
   \mathsf{state}_S(g)\notin\{\mathsf{FENCED},\mathsf{REVOKED}\}\ \wedge\\
  & A_g\cap\mathsf{Cut}_S(p(r))=\varnothing .
 \end{split}
\end{equation}
An ancestor may be fenced by completed succession without invalidating an
already-created descendant; inserting that ancestor's atom into
\(\mathsf{Cut}_S(p(r))\)
invalidates every dependent descendant because its complete closure contains
the atom.

A richer deployment may additionally bind per-atom observations such as
\(V_g^{\rm abs}=\{(a,E_S(a)):a\in A_g\}\) in a separately registered profile.
That vector is an optional refinement of the cut-set rule, not a field stored
or validated by the reference realization, and none of the results below
depends on it.

\begin{definition}[Active descendants]
\label{def:active-descendants}
The current active population below root \(r\) is
\[
 \mathsf{AD}_S(r)=\{g:\mathsf{root}(g)=r,\;
  \mathsf{state}_S(g)=\mathsf{ACTIVE},\;
  \mathsf{Current}_S(g)\}.
\]
A process whose record is fenced, revoked, or bound to a cut atom is not a
member and cannot pass fresh effect admission under A1 and A9.
\end{definition}

\subsection{Authority algebra, root grants, and population ceilings}

Let \((\mathcal Q,\preceq_Q,\vee,\bot)\) be the registered finite-join
semilattice of qualitative effect envelopes. The profile also supplies a
monotone membership predicate
\(\mathsf{Permits}_Q\subseteq\mathcal Q\times\mathcal E\): if
\(q\preceq_Q q'\) and \(\mathsf{Permits}_Q(q,e)\), then
\(\mathsf{Permits}_Q(q',e)\). Let \(D\) be the finite set of
quantitative resource dimensions and \(\mathcal K\subseteq\mathcal E\) the
effect classes for which concurrent holders are counted. A generation's
immutable granted allocation is
\begin{equation}
\label{eq:authority-allocation}
 \widehat\alpha_g=\langle q_g,\widehat b_g,u_g\rangle,
 \quad q_g\in\mathcal Q,\quad
 \widehat b_g\in\mathbb R_{\ge0}^{D},\quad
 u_g\in\mathbb N^{\mathcal K}.
\end{equation}
If \(\mathsf{Permits}_Q(q_g,e)\) for a counted class \(e\), profile validity requires
\(u_g(e)\ge1\). This prevents an idempotent lattice join from hiding the fact
that two siblings hold two concurrent copies of the same permission.

Let \(R_S(g)\) be the committed effect receipts attributed to \(g\) in state
\(S\), with every declared debit satisfying
\(\Delta b(\rho)\in\mathbb R_{\ge0}^{D}\). Its durable debit and remaining numeric authority are
\[
 d_g(S)=\sum_{\rho\in R_S(g)}\Delta b(\rho),\qquad
 b_g(S)=\widehat b_g-d_g(S),
\]
with well-formedness requiring \(0\le d_g(S)\le\widehat b_g\). The authority
currently transferable or exercisable by \(g\) is
\(\alpha_g(S)=\langle q_g,b_g(S),u_g\rangle\). For an authorization root
\(r\), define its irreversible lifetime debit
\[
 c_r(S)=\sum_{g:\mathsf{root}(g)=r}d_g(S).
\]
Fencing a generation therefore removes its residual allocation from the
active frontier but never removes its past debit from the root lifetime.

Define attenuation componentwise and define population composition by
\begin{equation}
\label{eq:authority-algebra}
 \begin{aligned}
 \alpha\sqsubseteq\alpha'&\iff
   q\preceq_Q q'\ \wedge\ b\le b'\ \wedge\ u\le u',\\
 \alpha\oplus\alpha'&=
   \langle q\vee q',\ b+b',\ u+u'\rangle .
 \end{aligned}
\end{equation}
Vector inequalities are pointwise. Let \(M\) be a finite labeled allocation
family whose entries have the form
\((g,\langle q,b,u\rangle)\), with unique generation labels. Write
\(\mathsf{lab}(M)\) for those labels and \(q_M(g)\), \(b_M(g)\), and \(u_M(g)\)
for the three components paired with label \(g\). Component aggregation uses
those projections, while the full labeled family is passed to the relational
predicate.

\begin{definition}[Root grant and population ceiling]
\label{def:population-ceiling}
Every authorization root \(r\) has an authenticated grant envelope
\[
 G_r=\langle q^G_r,b^G_r,u^G_r,h^G_r\rangle .
\]
Independently, every population \(p\) has a persistent ceiling
\[
 C_p=\langle q^P_p,b^P_p,u^P_p,\Phi_p,h^C_p\rangle ,
\]
where \(\Phi_p(M)\) is a registered downward-closed predicate over the full
labeled allocation family and \(\Phi_p(\varnothing)=\mathsf{true}\). A root
grant is admitted only if its component envelope is attenuated from the
corresponding population ceiling:
\[
 \langle q^G_r,b^G_r,u^G_r\rangle
 \sqsubseteq\langle q^P_{p(r)},b^P_{p(r)},u^P_{p(r)}\rangle .
\]
Define
\begin{equation}
\label{eq:root-ok}
 \begin{split}
 \mathsf{RootOK}_r(S,M)\iff{}&
  \bigvee_{g\in\mathsf{lab}(M)}q_M(g)\preceq_Q q^G_r\ \wedge
  c_r(S)+\sum_{g\in\mathsf{lab}(M)}b_M(g)\le b^G_r\ \wedge\\
 &\sum_{g\in\mathsf{lab}(M)}u_M(g)\le u^G_r,
 \end{split}
\end{equation}
and
\begin{equation}
\label{eq:population-ok}
 \begin{split}
 \mathsf{PopOK}_p(M)\iff{}&
  \bigvee_{g\in\mathsf{lab}(M)}q_M(g)\preceq_Q q^P_p\ \wedge
  \sum_{g\in\mathsf{lab}(M)}b_M(g)\le b^P_p\ \wedge\\
 &\sum_{g\in\mathsf{lab}(M)}u_M(g)\le u^P_p\ \wedge\Phi_p(M).
 \end{split}
\end{equation}
The combined exact-state guard is
\begin{equation}
\label{eq:safe-family}
 \mathsf{Safe}_r(S,M)\iff
 \mathsf{RootOK}_r(S,M)\wedge\mathsf{PopOK}_{p(r)}(M).
\end{equation}
The root conjunct conserves authority over one root lifetime, including
irreversible debit. The population conjunct limits current exposure even after
a new independently granted root becomes current. Its predicate can express
prohibited co-possession, separation of duties, branch-local target
constraints, and correlations not representable by fieldwise bounds.
\end{definition}

The reference realization instantiates this abstraction with finite sets of
registered effect labels, set membership for \(\mathsf{Permits}_Q\), and maps
of nonnegative safe-integer budgets. A bootstrap grant equals its population
ceiling; a later independent re-root grant may be strictly smaller and must be
componentwise bounded by that persistent ceiling. Root-lifetime validation
checks spent budget plus active remaining budget against the applicable root
grant. Separately, population validation prohibits the same effect label from
belonging to two concurrently active generations and checks their aggregate
remaining budgets against the population ceiling. Thus its effect-label
exclusivity realizes one holder unit per label. General semilattices,
arbitrary multiplicities, and arbitrary \(\Phi_p\) predicates describe the
registered abstraction and are not presented as executable-artifact results.

The ledger stores each generation's grant immutably. A budget-consuming effect
atomically appends one receipt/debit at commitment; remaining authority is
reconstructed from that grant and the receipts. Qualitative effects consume
no numeric budget unless their profile declares one, but the reference
realization still prohibits copying a label across active siblings.

\subsection{Certification, transfer, and re-rooting}

An evidence record for \(g\) has the canonical form
\[
 \zeta_g=\langle c,\kappa_c,\kappa_{\rm prop},\mathsf{gid}(g),m_g,
               h_{\rm evidence},\mathsf{APPROVE}\rangle .
\]
Let \(\mathsf{Ctl}_S(x)\) be the frozen registry mapping from a principal to
its admitted control root. The registered certification predicate is
\begin{equation}
\label{eq:cert-ok}
 \begin{split}
 \mathsf{CertOK}_S(g,\zeta_g)\iff{}&
   \mathsf{AuthenticatedCert}_S(\zeta_g)\ \wedge
   \kappa_c=\mathsf{Ctl}_S(c)\ \wedge\\
  &\kappa_{\rm prop}=\mathsf{Ctl}_S(\mathsf{proposer}(M_g))\ \wedge
   c\ne\mathsf{gid}(g)\ \wedge\\
  &\kappa_c\ne\kappa_{\rm prop}\ \wedge
   \zeta_g.\mathsf{gid}=\mathsf{gid}(g)\ \wedge\\
  &\zeta_g.\mathsf{manifest}=m_g\ \wedge
   \zeta_g.\mathsf{decision}=\mathsf{APPROVE}.
 \end{split}
\end{equation}
The first conjunct is the abstract external-authentication premise. The
reference artifact executes the remaining frozen-control-root and digest
binding checks; it does not execute caller authentication or public-key
signature verification.

For a single successor \(g\) of active predecessor \(p\), transfer requires
\begin{equation}
\label{eq:single-transfer}
 \mathsf{par}(g)=p\quad\wedge\quad
 \widehat\alpha_g\sqsubseteq\alpha_p(S).
\end{equation}
For an atomic fork with finite child set \(N\), \(|N|\ge2\), transfer requires
\begin{equation}
\label{eq:fork-transfer}
 \left(\forall g\in N:\mathsf{par}(g)=p\right)
 \quad\wedge\quad
 \bigoplus_{g\in N}\widehat\alpha_g\sqsubseteq\alpha_p(S).
\end{equation}
Every accepted family also satisfies the exact post-state root-grant and
population-ceiling bounds.
In the reference realization, \eqref{eq:fork-transfer} additionally means pairwise-disjoint child
effect-label sets and a componentwise sum of child budgets no greater than the
predecessor's reconstructed residual. No retained-predecessor allocation and
no fresh issuance participate in either core transition: \(p\) is fenced at
commit. A new authority source instead uses independent re-rooting.

Let \(\chi_{p,N}\) denote the authenticated authorization object for a family
transition. It binds the root, exact predecessor identifier and active state,
digest of \(\alpha_p(S)\), the child identifiers, manifests, allocations and
evidence records, the root-grant and population-ceiling commitments, the transition mode, and
every consumed identifier. For a staged singleton it also binds the compared
population version. Abstractly,
\begin{equation}
\label{eq:auth-ok}
 \begin{aligned}
 \mathsf{AuthOK}_S(p,N,\chi)\iff{}&
 \mathsf{Authenticated}_S(\chi)\ \wedge
 \mathsf{BindsExact}_S(\chi,p,N)\\
 &{}\wedge\mathsf{Fresh}_S(\chi)\ \wedge
 \bigwedge_{g\in N}\mathsf{CertOK}_S(g,\zeta_g).
 \end{aligned}
\end{equation}
The executable singleton path realizes \(\chi\) as a held reservation bound to
the predecessor residual and population version. The executable fork path
validates the complete family inline and records the atomic fork event; it
does not consume a collection of staged child reservations. In both paths,
the artifact executes the exact-binding and freshness guards in
\eqref{eq:auth-ok}; acceptance by an authenticated external caller remains the
A2 premise. No public-key authorization-record implementation is claimed.

\begin{definition}[Independent re-root]
\label{def:reroot}
Let \(r=\mathsf{curRoot}_S(p_0)\). An independent re-root of artifacts from
\(g\) into fresh root \(r'\ne r\), where \(p(r')=p_0\), is accepted only if
\[
 \{x:p(\mathsf{root}(x))=p_0\ \wedge\
       \mathsf{state}_S(x)=\mathsf{ACTIVE}\}=\varnothing .
\]
The transaction requires a fresh authenticated root grant whose issuer control
root is independent of the proposer control root, a strictly fresh root epoch,
independent accepted evidence, a fresh generation sequence,
\(\mathsf{par}(g')=\bot\), a new lineage closure containing no cut atom, and an
allocation bounded by the new grant and population ceiling. It checks
\(\mathsf{Safe}_{r'}\) on the exact post-state and atomically sets
\(\mathsf{curRoot}_{S'}(p_0)=r'\) while creating \(g'\) as the sole active
member of that population. Other populations and all historical root, debit,
cut, and generation records are unchanged. The source record may name \(g\),
but the new allocation and lineage derive only from \(r'\). Deleting an old
parent label or relabeling an old grant does not meet this definition.
\end{definition}

\begin{definition}[Rollback generation]
\label{def:rollback}
A rollback from current generation \(p\) to an earlier manifest \(M_a\) is the
ordinary successor transition with \(M_{g_{rb}}=M_a\),
\(\mathsf{par}(g_{rb})=p\), and a fresh population-issued sequence
\(n_{g_{rb}}\). It requires fresh evidence and authorization records; no record
belonging to the earlier generation \(a\) is reusable. Restoring the same
manifest through independent re-root instead creates a root generation \(g'\)
under Definition~\ref{def:reroot}; it is not the rollback generation
\(g_{rb}\) defined here.
\end{definition}

\subsection{Atomic succession and protected-effect admission}

The staged singleton path records a proposed candidate, independent evidence,
and a held reservation before activation. Quarantine is an execution
condition, not authority. The atomic fork path evaluates all children before
the commit and creates their active generation records together. In either
path, only \(\mathsf{ACTIVE}\) authorizes a fresh protected effect.

\begin{definition}[Atomic single-parent family handoff]
\label{def:handoff}
Let
\[
 h=\mathsf{CommitFamily}(p,N,\chi)
\]
be one serializable ledger transaction, where \(p\) is the unique active
predecessor and \(N\ne\varnothing\) is a finite set of distinct children. All
children have root \(r\), satisfy \(\mathsf{par}(g)=p\), and are not active
before the transaction. The guard re-resolves their identities, verifies that
\(p\) is current and active, evaluates \eqref{eq:cert-ok}--\eqref{eq:auth-ok}, rejects cut lineage atoms,
and checks \(\mathsf{Safe}_r\) on the exact post-state in which \(p\) is absent
and all members of \(N\) are active. For \(|N|=1\), \(\chi\) must be the held,
version-matched reservation. For \(|N|\ge2\), \(\chi\) is the complete inline
fork authorization bundle. At its single commit position \(\lambda_h\), the
transaction atomically consumes the applicable authorization object, marks
\(p\) \(\mathsf{FENCED}\), marks every \(g\in N\) \(\mathsf{ACTIVE}\), advances
the population version, and records the result. There is no durable state
between predecessor fencing and child activation.
\end{definition}

An ancestor cut \(\mathsf{RevokeAncestor}(p,a)\) atomically inserts \(a\) into
the monotone set \(\mathsf{Cut}_S(p)\). The ledger projection moves every nonterminal
affected generation and held reservation to \(\mathsf{REVOKED}\); already
terminal generations remain terminal. Safety does
not require finding or stopping every descendant process: a fresh effect or
population-changing request from any such process fails the current cut-set
check.

Let \(\mathsf{abs}:\mathcal O\rightarrow\mathcal E\) be the registered
operation normalizer, and let \(\lambda_o\) denote the admitted operation's
position in the ledger serial order. For operation \(o\in\mathcal O\) proposed by \(g\),
\(\mathsf{AdmitEffect}(g,o,id)\) is serialized with cuts and family handoffs.
If \(id\) already names the identical canonical effect request, the operation
returns the original receipt without a new admission, effect, or debit; this
remains a replay result even if \(g\) has since been fenced. Reuse of \(id\)
with a different canonical request is rejected. Otherwise the request is
fresh, and it is accepted only if \(g\in\mathsf{AD}_S(r)\), the exact normalized
effect \(e=\mathsf{abs}(o)\) satisfies \(\mathsf{Permits}_Q(q_g,e)\), and
sufficient budget remains.
A fresh acceptance atomically appends the declared debit, advances the
population version, and emits the digest-bound receipt
\begin{equation}
\label{eq:effect-receipt}
 \begin{split}
 \rho_o=H(\mathsf{Enc}(&d_{\rm receipt},id,r,\mathsf{gid}(g),H(A_g),\\
                        &H(\alpha_g(S)),H(o),e,\Delta b,\lambda_o)).
 \end{split}
\end{equation}
An attempt that fails any check may be logged, but it does not commit a
protected effect. Exact-result replay is specified only for effect requests;
other command classes rely on phase, uniqueness, version, and monotonicity
guards and receive no A10 retry guarantee.

\subsection{Safety invariants and main results}

\begin{definition}[Admissible initial state]
\label{def:initial-state}
An initial state satisfies \(\mathsf{InitOK}(S_0)\) when it is empty, or when
it contains a finite set of disjoint populations provisioned exclusively by
the bootstrap root-creation guard. For each provisioned population \(p\),
the state contains exactly one root record \(r\), exactly one corresponding
root-grant record \(G_r\), and exactly one generation record \(g\); there are
no other root, generation, reservation, authorization-consumption, cut, or
effect records for \(p\). The root is current and the generation is active:
\(p(r)=p\), \(\mathsf{root}(g)=r\), and
\(\mathsf{par}(g)=\bot\). The generation has a canonical fresh identity and
complete fresh root closure, and its profile-valid authenticated bootstrap
grant has an issuer control root distinct from its proposer control root. Its
lifetime debit is zero, its allocation satisfies
\(\mathsf{Safe}_r(S_0,\{(g,\alpha_g(S_0))\})\), and the population has no
effect receipt, consumed authorization object, held reservation, or cut atom.
Counters, root records, and current-root pointers are unique and mutually
consistent across the provisioned populations.
\end{definition}

For a root \(r\), abbreviate the exact current active allocation family as
\[
 \mathcal M_r(S)=\{(g,\alpha_g(S)):g\in\mathsf{AD}_S(r)\}.
\]

For every accepted-state prefix, define the following invariants:

\begin{description}
  \item[I1 (identity binding)] every registered generation record binds the
  canonical identity in \eqref{eq:generation-identity} to its resolved manifest and unique parent;
  \item[I2 (authorized activation)] every active generation satisfies its
  exclusive admission branch: a non-root generation has
  \(\mathsf{CertOK}\) evidence and was activated by one accepted
  \(\mathsf{AuthOK}\) family commit---the singleton path consumes its held
  reservation, whereas the fork path validates its complete inline bundle at
  the same commit that fences the predecessor; a bootstrap root generation has a profile-valid
  authenticated bootstrap grant whose issuer control root differs from the
  proposer control root; and an independent-re-root generation has a fresh
  authenticated root grant together with accepted independent evidence;
  \item[I3 (lineage closure)] every active generation record has its unique
  parent chain's complete atom closure, and that closure contains no cut atom;
  \item[I4 (allocation uniqueness)] a reservation, authorization-object
  identifier, generation identifier, effect-request identity, or budget debit
  has at most one consuming transition; exact effect replay is non-consuming;
  \item[I5 (root and population safety)] for every root \(r\),
  \(\mathsf{Safe}_r(S,\mathcal M_r(S))\);
  \item[I6 (attributable effects)] every protected-effect receipt binds one
  exact generation that was active and current at admission, an authorized
  effect, and one valid budget debit;
  \item[I7 (fence monotonicity)] a fenced generation and a cut lineage atom
  never become current again; resumed content enters through a fresh
  generation or an independent re-root.
\end{description}

\begin{theorem}[Population-Safe Succession]
\label{thm:population-safe}
Under A1--A10, I1--I7 hold after every prefix of every accepted trace that
starts from \(\mathsf{InitOK}(S_0)\). In particular, for every root \(r\) and
reachable state \(S\),
\begin{equation}
\label{eq:population-bound}
 \begin{aligned}
 \bigvee_{g\in\mathsf{AD}_S(r)}q_g&\preceq_Q q^G_r,&
 c_r(S)+\sum_{g\in\mathsf{AD}_S(r)} b_g(S)&\le b^G_r,\\
 \sum_{g\in\mathsf{AD}_S(r)}u_g&\le u^G_r,&
 \bigvee_{g\in\mathsf{AD}_S(r)}q_g&\preceq_Q q^P_{p(r)},\\
 \sum_{g\in\mathsf{AD}_S(r)}b_g(S)&\le b^P_{p(r)},&
 \sum_{g\in\mathsf{AD}_S(r)}u_g&\le u^P_{p(r)},\\
 \Phi_{p(r)}(\mathcal M_r(S))&=\mathsf{true}.&&
 \end{aligned}
\end{equation}
Every committed protected effect is attributable to exactly one generation
that was current and unrevoked at its admission position and lies within that
generation's allocation.
\end{theorem}

\begin{theorem}[Fork Conservation]
\label{thm:fork-conservation}
Consider an accepted atomic fork from predecessor \(p\) to children \(N\).
The composition of all child allocations is bounded by \(p\)'s reconstructed
pre-fork residual as in \eqref{eq:fork-transfer}, \(p\) is fenced at the same commit position, and
the residual can be consumed by that fork at most once. Repeating this argument
down a fork tree preserves the bound over current active leaf allocations
descended from its originating allocation; equal
qualitative permissions held by different descendants consume distinct holder
units under the abstract profile and are prohibited outright by the reference
effect-label profile.
\end{theorem}

\begin{theorem}[Cross-Generation Revocation Closure]
\label{thm:revocation-closure}
\leavevmode\par\noindent
Let \(c=\mathsf{RevokeAncestor}(p(r),a)\) insert \(a\) into
\(\mathsf{Cut}_S(p(r))\). No fresh
protected-effect admission ordered after \(c\) can be authorized by a
generation whose complete lineage set contains \(a\). This includes every
nested dependent descendant. A generation under an independent re-root is
outside the conclusion because its authority contains no old atom.
\end{theorem}

\begin{theorem}[Atomic Generation Handoff]
\label{thm:atomic-handoff}
For an accepted family handoff at \(\lambda_h\), a child in \(N\) has no
production authority before \(\lambda_h\), and predecessor \(p\) has no
production authority after \(\lambda_h\). Every concurrent fresh
protected-effect admission is ordered on exactly one side of \(\lambda_h\).
Hence no accepted prefix contains both \(p\)'s pre-transfer allocation and a
child's transferred allocation as simultaneously exercisable authority.
\end{theorem}

\begin{theorem}[Rollback Non-Reminting]
\label{thm:rollback}
If rollback generation \(g_{rb}\) has the same manifest digest as historical
generation \(a\), then \(\mathsf{gid}(g_{rb})\ne\mathsf{gid}(a)\). It cannot use
\(a\)'s evidence record, authorization object, reservation, activation record,
or cut lineage. Any authority obtained by \(g_{rb}\) comes from its fresh
accepted handoff and is counted by
\eqref{eq:population-bound}.
\end{theorem}

\begin{theorem}[No Self-Certification]
\label{thm:no-self-certification}
No trace consisting only of records originating in or controlled by a
non-bootstrap candidate generation \(g\) can move \(g\) to
\(\mathsf{ACTIVE}\). Every accepted candidate activation contains an accepted
evidence record from at least one profile-required control root independent of
the candidate proposer, plus an authenticated authorization decision of the
external lineage authority.
\end{theorem}

\begin{corollary}[Re-root non-laundering]
\label{cor:reroot}
An accepted independent re-root of \(g'\) has no authorization path to the old
root, imports no old-root allocation, and attributes every later admitted
effect solely to its fresh root. Removing or renaming old lineage fields
without the external root-grant transition cannot establish these facts.
\end{corollary}

The theorems establish authorization safety for the registered identity and
effect boundary. The abstract parameters above state the obligations for a
conforming profile; the artifact evidence in Section~\ref{sec:results} applies
only to the fixed manifest, authority, authentication, and transition
realization described here. Neither level claims semantic equivalence for
arbitrary code or models.

\section{Authorization Succession Protocol}
\label{sec:design}

The protocol turns mutation acceptance into a durable population transition.
The protocol does not ask a mutable agent to remember its own limits. It places
generation registration, allocation, succession, revocation, and protected
effect admission behind an external ledger and broker.

\subsection{Architecture}

\begin{figure*}[t]
\centering
\begin{tikzpicture}[
  font=\small,
  box/.style={draw, rounded corners=2pt, minimum height=8mm, align=center,
    inner xsep=5pt, inner ysep=3pt},
  blue/.style={box, draw=lfBlue, fill=lfBlueFill, text=lfBlue},
  purple/.style={box, draw=lfPurple, fill=lfPurpleFill, text=lfPurple},
  green/.style={box, draw=lfGreen, fill=lfGreenFill, text=lfGreen},
  amber/.style={box, draw=lfAmber, fill=lfAmberFill, text=lfAmber},
  arr/.style={-{Latex[length=2mm]}, line width=0.75pt},
]
\node[blue, minimum width=25mm] (runtime) at (0,1.8) {mutable candidate\\runtime};
\node[blue, minimum width=27mm] (identity) at (3.5,1.8) {canonical generation\\identity};
\node[purple, minimum width=28mm] (evidence) at (7.3,1.8) {independent evidence\\roots};
\node[amber, minimum width=25mm] (reserve) at (11.0,1.8) {exact succession\\authorization};

\node[green, minimum width=82mm, minimum height=10mm] (ledger) at (6.4,0)
  {external generation registry + population ledger\\
   atomic replacement/fork, lineage cuts, effect-request replay index};

\node[purple, minimum width=29mm] (active) at (2.6,-2.0) {durable active\\frontier};
\node[green, minimum width=29mm] (broker) at (6.4,-2.0) {protected-effect\\broker};
\node[blue, minimum width=29mm] (effect) at (10.2,-2.0) {typed committed\\effect};

\draw[arr, lfBlue] (runtime) -- (identity);
\draw[arr, lfBlue] (identity) -- (evidence);
\draw[arr, lfPurple] (evidence) -- (reserve);
\draw[arr, lfAmber] (reserve) -- (reserve |- ledger.north);
\draw[arr, lfGreen] (ledger.south -| active) -- (active.north);
\draw[arr, lfPurple] (active) -- (broker);
\draw[arr, lfGreen] (broker) -- (effect);
\draw[arr, lfRed] (broker.north) -- node[right, text=lfRed, fill=white,
  inner sep=1.5pt] {current-state check} (broker.north |- ledger.south);
\end{tikzpicture}
\caption{Authorization-succession architecture. Candidate production and evidence feed
an external ledger; only a committed population transition changes the active
generation frontier. The effect broker revalidates current lineage and
allocation state at the protected commitment point.}
\Description{A top row contains mutable candidate runtime, canonical
generation identity, independent evidence roots, and exact succession
authorization. They feed an external generation registry and population ledger.
The ledger determines the active generation frontier, which is checked by a
protected-effect broker before a typed effect commits.}
\label{fig:lf-architecture}
\end{figure*}
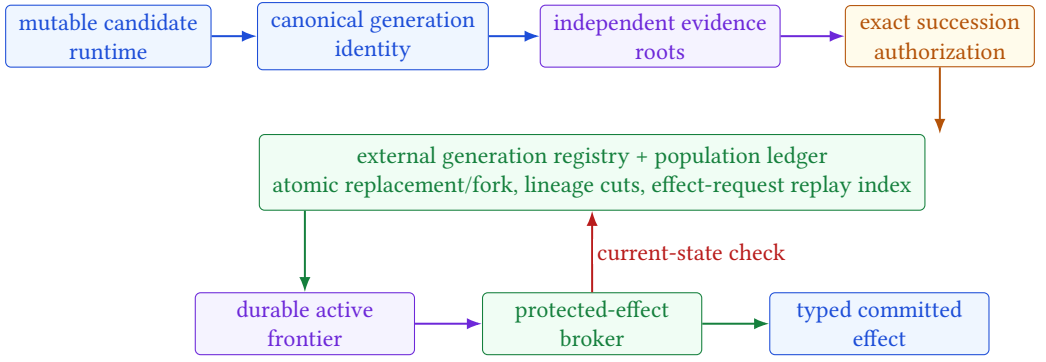

Figure~\ref{fig:lf-architecture} separates candidate production from authority
control. Mutable runtimes may produce any number of candidates. The identity
builder commits every security-relevant field and the evidence layer evaluates
registered predicates. Neither component can activate a candidate. Only the
population ledger can accept the applicable authorization object and
atomically change the active frontier. Every protected effect then revalidates
the resulting generation and allocation state at the broker.

\subsection{Canonical generation identity}

The registry computes a generation identifier only from a complete canonical
record. Table~\ref{tab:generation-identity} separates the executable reference
record from profile-level extensions. Omission or addition of a field in the
fixed reference schema is a registration error rather than an implicit
wildcard.

\begin{table*}[t]
\caption{Generation identity inputs and their realization status.}
\label{tab:generation-identity}
\small
\begin{tabularx}{\textwidth}{L{0.18\textwidth}Y L{0.27\textwidth}}
\toprule
Layer & Bound fact & Security role \\
\midrule
Reference manifest & code digest, control path, enforcement-configuration digest, memory-policy digest, model-route digest, policy-compiler digest, proposer identifier, Skill-closure digest & fixes the exact eight-field executable schema \\
Reference lineage & population and root identifiers, root epoch and grant digest, one parent-generation identifier, complete lineage-atom set & makes single-parent succession and fork ancestry explicit \\
Reference lifetime & fresh population-issued generation sequence & separates launches and blocks historical identity replay \\
Reference evidence & proposer and certifier principals mapped through frozen control roots; subject-manifest and evidence digests & checks digest binding and declared-root separation \\
Registered extension & any additional prompt, tool, harness, dependency, normalizer, or profile field named by a larger governed schema & extends the abstract identity obligation only when a deployment resolves and binds it \\
\bottomrule
\end{tabularx}
\end{table*}

The identifier is the domain-separated digest in
\eqref{eq:generation-identity}, over the exact manifest
digest, population, root grant, generation sequence, unique parent, and complete
lineage-atom set. Domain separation prevents a manifest digest, authorization
record, or effect receipt from being reinterpreted as a generation identity.
The fresh generation sequence makes two otherwise identical manifests distinct.
A rollback may intentionally reuse a manifest commitment while receiving a new
sequence and therefore a new \(\mathit{gid}\). Authenticated software provenance
and rollback-resistant update metadata supply compatible lower-layer
techniques \cite{intoto2019,tuf2010}; the generation record adds the live
authorization lifetime. Claims backed by the executable artifact apply to the
eight-field reference manifest, not automatically to the extension row.

\subsection{Candidate and evidence phases}

The staged single-successor path records a \mcode{PROPOSED} generation,
attaches an independent certification record, moves it to \mcode{RESERVED},
and atomically changes it to \mcode{ACTIVE} while fencing its predecessor.
Quarantine describes where candidate code is evaluated; it is not an
authority-bearing ledger phase. The fork path instead evaluates a complete
child family and, after inline validation, creates all child
\mcode{ACTIVE} records in the same transaction that fences the predecessor.
Terminal \mcode{FENCED} and \mcode{REVOKED} states are monotone. A changed
manifest creates a new candidate or fork-child description rather than
editing an accepted record.

At the registered-profile level, evidence slots may denote typed predicates
such as body-manifest integrity, transition-envelope compliance, test-policy
satisfaction, and human or organizational approval. Such a profile names its
admitted principals and required independent control-root classes. The
reference realization implements the narrower evidence-record boundary used
by the experiments: one opaque evidence digest and \mcode{APPROVE} decision are
bound to the exact generation and subject manifest, and the proposer and
certifier identifiers are resolved through a frozen declared-control-root
registry and required to differ at that root. It does not evaluate the opaque
evidence, implement a multi-slot predicate policy, authenticate the caller,
prove key possession, or verify a public-key signature. Those obligations and
the substantive evidence judgment remain A2 and profile-level premises.
A staged authorization is also bound to the predecessor's
residual and population version, so reuse against a later allocation state
fails.

\subsection{Exact authority authorization}

For staged single-successor promotion, the ledger reserves the proposed
allocation against one population version and one digest of the unique
predecessor's current residual. The reservation binds the complete effect-label
set, additive budgets, certification record, predecessor, child, and single-use
reservation identity. Creating it does not authorize effects. In the reference
realization, a held reservation rejects new predecessor effects; any other
accepted population transition advances the compared version. Either guard
prevents concurrent consumption from spending the reserved residual.

An atomic fork does not reuse that staged-reservation protocol. Its command
supplies at least two complete child descriptions, and the ledger validates
their manifests, independent evidence, allocations, common predecessor,
disjoint effect labels, summed budgets, and exact post-state inline. It then
fences the predecessor and creates all children in one commit. Both paths may
leave residual authority unallocated; neither retains an attenuated active
predecessor, returns spent budget, or mutates the predecessor's immutable
grant. The ledger derives transferable residual from the immutable grant minus
committed receipts. Across an authorization root, cumulative lifetime debit
plus every active generation's residual remains at or below that root's grant.
Separately, the aggregate current active residual remains at or below the
population's persistent ceiling, including after an independent re-root.

\subsection{Atomic single-parent family handoff}

Algorithm~\ref{alg:commit-replacement} gives the common safety shape. The
transaction starts from one current predecessor, reconstructs its residual and
lineage, verifies the mode-specific authorization object, and writes the
predecessor fence and child activations in one state change. The exact
post-state, not an intended sequence of remote messages, defines the
linearization point.

\begin{algorithm}[t]
\caption{Atomic single-parent family commit}
\label{alg:commit-replacement}
\begin{algorithmic}[1]
\Require predecessor $p$, nonempty child set $N$, authorization object $\chi$
\State set $r\gets\mathsf{root}(p)$; begin serializable transaction at population version $v$
\State require $p$ is current, active, and has no cut lineage atom
\State require every $g\in N$ has unique parent $p$ and a fresh identity
\State require every child manifest and evidence record passes its exact checks
\If{$|N|=1$}
  \State require $\chi$ is the held reservation bound to $(v,p,N)$ and the current residual
\Else
  \State require $\chi$ is a complete inline fork bundle and $|N|\ge2$
\EndIf
\State require \eqref{eq:single-transfer} or \eqref{eq:fork-transfer}, as applicable
\State compute the exact post-state labeled allocation family $M'$ excluding $p$ and including $N$
\State require $\mathsf{Safe}_r(S',M')$
\State atomically write $p\mapsto\Fenced$, $N\mapsto\Active$, consume $\chi$
       where applicable, advance $v$, and persist the result
\State commit
\end{algorithmic}
\end{algorithm}

The active frontier is the current, effect-admissible set in
Definition~\ref{def:active-descendants}; a cut atom or terminal state removes a
generation from that set. The broker serializes its admission against all
population transitions and reads the current ledger version and state. Agent transaction and provider-lifecycle systems already
supply compatible staged effects and retry-safe lower boundaries
\cite{cordon2026,aidguard2026}; this transaction changes which software
generations hold the upstream authority. A fresh effect admission belongs either before the family commit and to the
active predecessor allocation, or after it and to an active child allocation.
It cannot observe a committed state in which both transferred allocations are
active. A crash before commit leaves the predecessor unchanged and any staged
candidate non-active; a crash after commit recovers the complete post-state.

\subsection{Fork and nested succession}

A fork atomically replaces one predecessor with at least two children. Every
child independently names that same predecessor as its unique authorization
parent; the predecessor never remains active after a successful fork. Additive
budgets are partitioned so their componentwise sum does not exceed the
predecessor's reconstructed residual. The reference profile allocates effect
labels disjointly across children, so no protected effect is silently copied.
The abstract algebra in Section~\ref{sec:formal} permits a registered
multiplicity-aware relation, but the reference artifact does not implement a
more general shareability rule.

Nested succession repeats the same single-parent rule against the current
population state. An intermediate generation cannot launder authority by
allocating to children and then disappearing: each fork consumes one current
predecessor residual exactly once, fences that predecessor, preserves all past
debits, and introduces only a bounded child composition. The core protocol has
no multi-parent merge transition. Combining independently authorized branches
requires a separately governed construction outside the core theorem rather
than an implicit union of their authority.

\subsection{Ancestor cuts and independent re-root}

Revoking an ancestor inserts its atom into the population's monotone cut set
and marks nonterminal affected generations and held reservations revoked;
already terminal generations remain terminal.
Activation, succession, fork, rollback, and every fresh effect admission reject
a witness containing the cut atom. Offline descendants need not receive a push
message for safety; their next protected transition encounters the current
cut. Frontier accounting is required before a positive statement that every
dependent descendant has become terminal or re-rooted.

An independent re-root transaction creates a fresh root and allocation from an
authenticated external grant record. It binds the current manifest, new root
epoch and lineage atom, effect labels, budgets, and independent certifier
control root. The new generation has no authorization
parent; the old generation is already non-active before re-root admission.
Re-root therefore introduces a distinct external authority cause rather than
deleting or relabeling the old atom. The formal model permits a separate,
non-authorizing source record for provenance, and the external adapter evidence
uses source-native identifiers for audit correlation; the reference re-root
authorization record itself contains no provenance/source field. The
population-wide cut set and persistent ceiling remain in force across the
current-root switch.

\subsection{Rollback without reminting}

Rollback begins with ordinary candidate creation from the selected historical
manifest. The ledger assigns a fresh generation sequence and binds the unique
current predecessor. Optional mutation metadata can preserve a non-authorizing
audit correlation, but the reference authorization decision does not depend on
a provenance field. Fresh evidence and authorization records are required.
Historical activation records, consumed reservations, cut atoms, and spent
budgets cannot be reused.
Thus a known-good body can be restored promptly without treating old authority
as a property of its bytes. Checkpoint replay and safe-restore work already
establish the general authority-resurrection threat
\cite{acrfence2026,agentlibos2026}; the rule here specializes it to a population
transfer whose identity, allocation, and authorization record are fresh.

\subsection{Protected-effect admission}

For request $x$, the broker atomically verifies:

\begin{enumerate}
  \item the exact generation record authenticates and belongs to the governed population;
  \item the generation is active in the serialized ledger state, whose transition invariant maintains one current authorization root;
  \item no ancestor atom in its authority witness is cut or stale;
  \item the complete normalized effect belongs to its allocated relation;
  \item declared additive budget remains available in both the generation
  allocation and root-lifetime accounting; and
  \item the request identity has not committed a different canonical effect.
\end{enumerate}

For a fresh request identity, acceptance consumes required resources, advances
the population version, and emits a typed receipt in the same transaction. An
exact retry returns that receipt without a second debit; the same identity
with a different canonical request is a conflict. A stale process cannot
authorize from a cached allocation. Missing current facts remain
non-authorizing.

\subsection{Evidence and independent verification}

Every accepted command emits a hash-linked, operation-specific event. Its
envelope binds the previous digest and event type; its payload binds the
population, generation, and transition fields used by that operation. Effect
authorization events additionally bind the effect-request identity and
canonical request digest. The case transcripts and final result-set commitment
separately bind registered inputs, decisions, terminal state, and the complete
denominator. The independent checker rebuilds
identity digests, the single-parent generation forest, active frontiers, allocations, cuts,
budget conservation, phase monotonicity, evidence-root independence, and
effect decisions without importing the controller or adapter implementation.

Semantic regression tests modify a security-relevant fact and then consistently
reseal the enclosing event, transcript, state, case, and suite records. Each
regression declares which content-addressed inner objects are also rebuilt.
The checker must reject the resulting trace for the registered invariant or
denominator violation, rather than only because an unrelated outer checksum is
stale.

The checker and executable cases validate the fixed eight-field manifest,
frozen control-root registry, digest-bound records, single-parent succession,
atomic fork, disjoint effect labels, and integer budgets described above.
General profile digests, extra manifest fields, per-atom epoch vectors,
public-key signatures, arbitrary relational authority predicates, and
multi-parent merge are outside that executable evidence claim.

\section{Protocol Instantiations}
\label{sec:instantiations}

The executable artifact first uses two separately implemented logical runtime
mappings to test whether the succession contract depends on one command
layout. They are provider-free mappings, not measurements of named external
systems. Section~\ref{sec:evidence} reports a separate refinement through
adapters to two frozen public runtimes.

\subsection{Direct-call mapping}

The \mcode{logical-direct-call/v1} adapter receives a typed action and payload,
maps the action through an explicit action-to-operation table, and invokes the
ledger in the caller's request path. Its transcript binds the complete input,
mapped operation, ledger event sequence, decision, and preceding transcript
digest. This shape makes the normalized command boundary directly visible.

\subsection{Message-mailbox mapping}

The \mcode{logical-mailbox/v1} adapter first translates the same typed action
into a versioned topic and delivery envelope. A topic-specific handler then
constructs the ledger command. The delivery ordinal is transport metadata and
cannot change the semantic request. Its separate topic and handler tables
exercise a message-oriented dispatch shape without presenting an external
provider as evaluated.

\subsection{Common mapping contract}

Table~\ref{tab:adapter-contract} lists the common contract facts established by
the adapter-to-ledger path; generation freshness is assigned inside the ledger
rather than supplied by either adapter. They share the typed input contract,
transcript recorder, and semantic ledger, while their dispatch translations are separately coded. The
experiment therefore tests translation-shape agreement; it does not treat the
two paths as independent semantic implementations.

The frozen executable profile instantiates one authorization parent per
successor. It covers replacement, predecessor-fencing fork allocation, nested
succession, rollback, and independent re-root. It has no multi-parent merge
transition; branch coordination can authorize a complete effect only through
one current witness or a separately issued root.

\begin{table*}[t]
\caption{Logical runtime-to-succession mapping contract.}
\label{tab:adapter-contract}
\small
\begin{tabularx}{\textwidth}{L{0.22\textwidth}Y Y}
\toprule
Contract field & Direct-call source & Message-mailbox source \\
\midrule
Authorization parent & typed predecessor identifier & topic-handler predecessor identifier \\
Body commitment & canonical manifest fields & canonical manifest fields \\
Generation freshness & ledger-owned sequence assigned during the command & ledger-owned sequence assigned while handling the message \\
Evidence & direct certification command & certification topic and handler \\
Requested authority & direct reservation command & reservation topic and handler \\
Lifecycle signal & explicit succession operation & versioned succession message \\
Protected effect & direct normalized request & effect-attempt message \\
\bottomrule
\end{tabularx}
\end{table*}

For both mappings, native success is evidence rather than authority. An
evaluation score, test pass, selected archive status, successful repository
creation, or launched process cannot bypass the external activation receipt.
The contract also carries source-native identifiers for audit correlation, but
effect authorization receives the canonical generation identifier and loads
the lineage, allocation, reservations, and population state inside the
serialized ledger operation. The ledger transition invariant, not a supplied
version, maintains current-root consistency.

\subsection{Mutation-surface normalization}

The registered corpus changes six security-relevant surfaces: code, model
route, Skill/dependency closure, memory policy, policy compiler, and
enforcement configuration. Each adapter maps its native representation into
the same typed manifest. Four outcomes are then distinguished per surface:

\begin{itemize}
  \item a safe refactor preserves the complete effect relation and control
  path;
  \item a safe narrowing produces a strict sub-allocation;
  \item an effect expansion requests at least one tuple or quantitative share
  outside the available parent/population authority; and
  \item a control-path weakening removes or bypasses a required identity,
  evidence, ledger, or broker check.
\end{itemize}

The evaluation is not a general semantic-equivalence classifier. Corpus
records provide the normalized ground-truth delta. The experiment asks whether
both mappings preserve that delta and whether the ledger enforces its exact
succession consequence.

\subsection{Durability mapping}

The durable experiment stores the complete ledger image behind an intent record
and temporary next-state image. A \mcode{COMMIT\_SUCCESSION} execution first
constructs and validates the complete post-state, persists the intent and next
image, atomically renames that image, fsyncs the containing directory, and
removes the intent. Restart verifies the event chain and accepts only the base
or intended next-state digest.

The 16 registered crash cuts are persistence-level boundaries around opening,
writing, fsyncing, closing, renaming, and cleaning those three files. They
exercise the atomic predecessor-fence/successor-activation publication path;
they are not presented as separate crash experiments for every protocol
operation. At every cut, recovery must expose exactly one active generation and
remove intent and temporary residue.

\subsection{External-runtime refinement contract}

A concrete external adapter is admitted only when it binds every field in
Table~\ref{tab:adapter-contract} to a frozen public version and identifies the
actual launch and protected-effect boundaries. The evidence record must include
repository URL, license, immutable commit, acquisition date, executed command,
native lineage fields, mapping code digest, and produced trace digest. At least
one path must exercise a real body change, descendant or candidate creation,
process restart, and a predecessor/successor overlap attempt.

Public systems that implement versioned bodies, recursive descent, or
self-modifying archives provide strong substrate candidates
\cite{ourark2026,dgm2025}. Their native identity and evolution mechanisms are
treated as inputs. The evaluated contribution remains the population
authorization transition placed around those mechanisms.

\section{Evaluation}
\label{sec:evaluation}

The evaluation is organized as separately gated evidence layers. The first
layer executes a fixed mutation-and-lineage corpus against the reference
ledger. The second maps the same cases through direct-call and message-mailbox
dispatch paths. The third interrupts durable replacement publication and
enumerates registered contender orders. The fourth asks an independent checker
to reconstruct the invariants and reject consistently resealed semantic
mutations. Two concrete refinements over frozen public runtimes have their own source,
execution, and result seals; logical adapter results are never relabeled as
external-runtime evidence.

\subsection{Research questions}

\begin{description}[leftmargin=3.1em,style=nextline]
  \item[RQ1: Separation.] Do individually acceptable sibling transitions,
  cached descendant credentials, non-atomic promotion, and historical rollback
  produce effects under adjacent per-session or per-generation controls while
  the population protocol rejects the same executions?

  \item[RQ2: Precision.] Does the protocol preserve every registered safe
  refactor, safe narrowing, legal single-successor promotion, and independent
  re-root while rejecting expansion and control-path weakening?

  \item[RQ3: Conservation.] Across the registered fork, nested-succession,
  ancestor-cut, effect-consumption, and rollback profiles, do complete effect
  relations remain within the persistent population ceiling, and does root-lifetime
  debit plus every active residual remain within the original root grant?

  \item[RQ4: Atomicity and recovery.] Does durable replacement handoff recover
  to a complete pre-state or post-state at every registered persistence cut,
  and do the registered contender orders preserve one winner and reject the
  stale predecessor?

  \item[RQ5: Independent verifiability.] Can a separately implemented checker
  reconstruct all registered decisions and reject security-relevant traces
  after an adversary modifies semantics and consistently recomputes hashes?
\end{description}

\subsection{Registered 32-case corpus}

The corpus is fixed before interpreting results. Twenty-four cases cross six
mutation surfaces with four outcomes. The remaining eight exercise generation
and population transitions. Table~\ref{tab:registered-corpus} gives the complete
logical denominator.

Every registered successor in this corpus has one authorization parent. The
corpus therefore evaluates the paper's single-parent authorization-lineage
model; it contains no multi-parent merge transition.

\begin{table*}[t]
\caption{Registered 32-case logical corpus.}
\label{tab:registered-corpus}
\footnotesize
\begin{tabularx}{\textwidth}{L{0.23\textwidth}L{0.28\textwidth}L{0.15\textwidth}Y}
\toprule
Group & Cases & Expected & Distinguishing obligation \\
\midrule
Mutation surfaces & code, model route, Skill/dependency closure, memory policy, policy compiler, enforcement configuration $\times$ safe refactor & 6 \Allow{} & exact effect relation and control path preserved \\
Mutation surfaces & same six surfaces $\times$ safe narrowing & 6 \Allow{} & strict sub-allocation with no returned spent resource \\
Mutation surfaces & same six surfaces $\times$ effect expansion & 6 \Deny{} & requested tuple/share exceeds parent or population residual \\
Mutation surfaces & same six surfaces $\times$ control-path weakening & 6 \Deny{} & identity, evidence, ledger, or broker mediation removed \\
Legal succession & single replacement; independent re-root & 2 \Allow{} & atomic old/new handoff; old-lineage dependence explicitly severed \\
Population attacks & sibling split; nested-successor laundering & 2 \Deny{} & aggregate allocation cannot be copied or hidden by depth \\
Lineage attacks & ancestor-revoke leak; rollback replay & 2 \Deny{} & cut atom and terminal authorization remain non-current \\
Integrity attacks & self-certification; old/new concurrent & 2 \Deny{} & independent control roots and one atomic active frontier required \\
\bottomrule
\end{tabularx}
\end{table*}

The expected total is 14 allows and 18 denies. Each case specifies the initial
root grant and population ceiling, registered generations and parents, pre-state allocations,
candidate manifest delta, evidence roots, operation sequence, complete effect
attempts, and expected typed result. Unknown fields are explicit and cannot be
filled from adapter defaults.

\subsection{Two logical runtime mappings}

Both adapters execute all 32 cases, producing 64 paired mappings. The
direct-call path uses an explicit action-to-operation table. The mailbox path
uses versioned topics, delivery envelopes, and topic-specific command handlers.
Both submit the same canonical succession contract to the same ledger and must
produce the same case verdict and effect disposition. They share the input
contract and transcript recorder; their independent variable is dispatch
shape, not policy semantics.

Agreement is necessary but not treated as independent ground truth. The
corpus registration supplies expected outcomes, while the checker separately
reconstructs whether each trace satisfies the formal invariants. A mapping-
substitution regression changes one adapter transcript without altering the
other; a cross-adapter regression changes one durable post-state. The checker
must reject both disagreement forms.

\subsection{Logical baselines}

The baseline layer implements four deliberately minimal stated guarantees and
labels each result as a logical abstraction rather than execution of an
authors' artifact:

\begin{enumerate}
  \item \emph{session continuity} accepts an effect while the presenting
  session remains live and its local grant has not expired;
  \item \emph{exact-hash all-deny} requires the candidate body and runtime
  manifest to equal the predecessor exactly;
  \item \emph{per-generation fixed ceiling} checks each accepted successor
  against the parent's transition envelope and ceiling independently;
  \item \emph{delegation-edge revocation} rejects a credential on a revoked
  registered delegation path but has no generation replacement or rollback
  transaction.
\end{enumerate}

External only-path enforcement is held constant as a deployment premise: every
protected effect reaches the configured broker. It is not counted as a fifth
baseline because mediation alone supplies no population predicate to classify.

The strict gate requires two complementary observations. A local continuity or
per-generation ceiling must accept at least one registered population
violation, demonstrating the need for joint state. Exact-hash all-deny must
reject at least one registered safe successor, demonstrating that universal
reauthorization is not the only safe policy. The reference mechanism must preserve
all registered benign cases and reject all registered violations.

\subsection{Durable crash cuts and interleavings}

Crash injection targets 16 named persistence boundaries in one durable
handoff: before and after opening, writing, fsyncing, and closing the intent
and next-state files; before and after atomic rename; and around directory
fsync and intent cleanup. Each worker terminates in a separate process.
Recovery starts from disk, validates the hash-linked ledger, removes residue,
and must expose either the active predecessor plus reserved successor or the
fenced predecessor plus active successor---never a dual-active or zero-active
replacement state.

The schedule suite enumerates all 32 values of five Boolean order choices for
two competing successors: proposal order, certification order, reservation
order, whether the losing or winning commit is attempted first, and whether
the active-successor or stale-predecessor effect probe runs first after
handoff. Calls within one schedule are serialized. These registered order
vectors are not a claim over every concurrent distributed execution or over
cut/rollback interleavings; they exercise unique reservation, one successful
handoff, losing-commit rejection, and stale-predecessor effect rejection.

\subsection{Independent checker and semantic mutants}

The runner serializes each event as a canonical hash-linked record and binds
the complete result set. The checker imports no ledger, controller, adapter, or
corpus expected-verdict function. It holds a separately encoded case oracle,
recomputes domain-separated digests, reconstructs the generation graph and
active frontier, and validates phase monotonicity, evidence-root independence,
population joins, the bijection between effect events and durable
consumptions, root-lifetime budget conservation, published population
summaries, cuts, exact effect-request retries, and effect receipts.

The adversarial regression harness changes a security-relevant fact and then
recomputes the enclosing integrity layers. Registered regressions cover
identity and lineage substitution, missing ancestry, active revoked lineage,
population ceiling breaches, illegal phase transitions, certificate/state
mismatch, control-root aliasing, reservation reopening, verdict and oracle
flips, accepted fenced effects, incomplete revocation, false re-root, rollback
identity reuse, fabricated or detached consumptions, corrupted root-budget
summaries, cross-adapter divergence, denominator omission or duplication, and
mapping substitution. A regression is counted only when the checker
rejects the resealed input for the intended semantic or denominator reason.

\subsection{Metrics and decision gate}

Primary safety metrics are unexpected allows, population-envelope violations,
post-cut descendant effects, dual-active replacement effects, historical grant
remints, self-certified activations, budget imbalance, replay acceptance, and
mutant escape. Precision metrics are safe-successor retention and the number of
benign cases forced to obtain an entirely new root. Evidence metrics are case,
adapter, crash-cut, schedule, and mutant numerators over their fixed
denominators. Runtime measurements, when reported, are descriptive diagnostics
for the local artifact.

The executable gate requires 32/32 logical outcomes, 64/64 adapter mappings,
zero registered safety violations under the executed reference profile, all crash cuts and
schedules in a complete pre/post state, checker agreement on every original
trace, and rejection of every registered consistently resealed regression.
Any failed or unexecuted
case remains visible in its denominator.

\subsection{External refinement gate}

The external layer freezes two public self-modifying or self-evolving runtime
versions. Each adapter must bind native identity, lineage, validation, launch,
and effect observations to the succession contract and publish its repository,
license, immutable revision, mapping fields, commands, and result seal. Across
the layer, at least one execution must perform a real body edit or descent,
restart the descendant in a fresh process, complete a broker-mediated handoff,
and reject an old-instance effect after the fence. Each runtime then maps the
registered 32-case corpus without treating those replays as independent
evolution runs. Ground truth comes from immutable source history, native
validation output, the succession ledger and broker, and the registered case
oracle. The separately published harness-tampering taxonomy informs surface
selection \cite{harnesstampering2026}; its annotated corpus is not part of the
executed evidence reported here.

\section{Evidence Status and Results}
\label{sec:evidence}
\label{sec:results}

All observations in this section are exact finite-corpus results regenerated
from the frozen inputs.  We maintain three separately sealed evidence bundles: the
provider-free core artifact, the OurArk adapter, and the Darwin G\"odel Machine
(DGM) adapter.  Each bundle has its own aggregate seal and verifier.  The core
seal does not cover either public checkout or external result; conversely, an
external seal does not enlarge the core denominator.  This separation makes a
missing or changed external checkout incapable of silently changing the short
core gate.

\subsection{Core conformance, faults, and adversarial checking}

Table~\ref{tab:core-results} reports the complete core denominator.  Each of
the 32 logical cases was executed once through the direct-call mapping and
once through the mailbox mapping.  Both mappings therefore produced 14
allows and 18 denies, for 28 allows and 36 denies across 64 replays, with no
verdict or effect-disposition disagreement.  The independently implemented
checker accepted all 64 original traces and rejected all 28 registered
outer-resealed semantic mutants.  ``Outer-resealed'' means that the harness
recomputed the enclosing event, transcript, summary, state, and case
commitments after changing a security-relevant fact; it does not claim an
exhaustive mutation space.

\begin{table*}[t]
\caption{Frozen core evidence.  A pass count is over the complete registered
denominator shown in the second column.}
\label{tab:core-results}
\small
\begin{tabularx}{\textwidth}{L{0.25\textwidth}L{0.25\textwidth}Y}
\toprule
Evidence family & Frozen denominator & Result and interpretation \\
\midrule
Logical corpus & 32 cases; 14 \Allow{}, 18 \Deny{} & 32/32 matched the registered verdict under the reference mechanism \\
Command-surface mappings & 32 direct-call + 32 mailbox replays & 64/64 matched; 28 allows, 36 denies; zero cross-mapping disagreement \\
Independent checking & 64 complete traces & 64/64 accepted after independent reconstruction \\
Semantic regressions & 28 outer-resealed mutants & 28/28 rejected for the registered semantic or denominator violation \\
Reference-profile invariants & 12 targeted probes & 12/12 passed: fork allocation, cut, rollback, lifetime debit, idempotency, reservation, and re-root \\
Durable handoff cuts & 16 process terminations & 12 recovered the complete pre-commit state and 4 the complete post-commit state \\
Contender schedules & 32 deterministic command orders & 32/32 preserved one winner and rejected the stale predecessor \\
\bottomrule
\end{tabularx}
\end{table*}

The 12 profile probes supply focused evidence for RQ3 beyond case-level
classification.  In the succession debit probe, 90 consumed tokens remain
charged to the authorization root, the successor can consume only the final
10, and a further debit is rejected.  The fork probe similarly records 25
spent tokens plus 75 units of active residual against a 100-token root grant.
Rollback cannot restore a predecessor's spent allocation; an exact effect
retry returns the original effect identifier bound to its receipt without a
second debit; a conflicting body
under the same request identifier is rejected.  While a handoff reservation
is held, a new predecessor debit is rejected, while an exact retry of the
already committed request remains idempotent.  These observations exercise
the root-lifetime accounting and predecessor-residual/version binding used by the formal
invariants rather than merely checking terminal labels.

The crash experiment terminates a worker at 16 named file-persistence cuts
around one succession commit.  Every cut recovered to one of the two complete
states; none exposed a committed dual-active or zero-active replacement.  The
schedule result covers all combinations of the five registered order bits,
but remains a sequential deterministic command experiment rather than a
claim about arbitrary thread, filesystem, or distributed executions.

\subsection{Controlled baseline separation}

Table~\ref{tab:baseline-results} compares only the four stated logical
abstractions from Section~\ref{sec:evaluation}; these rows are not executions
of third-party systems.  Session continuity preserves every registered benign
case but also accepts all 18 registered violations.  Exact-hash all-deny gets
the 18 deny cases right while rejecting every benign successor.  The
per-generation ceiling misses sibling copying and two inherited-lineage
violations.  Delegation-edge cascade misses nine cases involving
control-path weakening, sibling partition, handoff overlap, or rollback.
The reference mechanism is the only row that classifies all 32 cases as registered.

\begin{table}[t]
\caption{Logical classifications matching the registered oracle.}
\label{tab:baseline-results}
\small
\begin{tabularx}{\columnwidth}{Y r}
\toprule
Mechanism or stated abstraction & Match \\
\midrule
Session continuity & 14/32 \\
Exact-hash all-deny & 18/32 \\
Per-generation fixed ceiling & 29/32 \\
Delegation-edge cascade & 23/32 \\
Complete authorization-succession profile & 32/32 \\
\bottomrule
\end{tabularx}
\end{table}

\subsection{What the reference realization implements}

The formal core retains single-parent authorization lineage while abstracting
over the identity closure, authority algebra, and authenticated-record
predicate.  The executable evidence concerns the bounded realization in
Table~\ref{tab:reference-profile}; it does not silently substitute one concrete
commitment mechanism for every admissible deployment realization.

\begin{table*}[t]
\caption{Reference realization boundary.}
\label{tab:reference-profile}
\small
\begin{tabularx}{\textwidth}{L{0.21\textwidth}L{0.36\textwidth}Y}
\toprule
Dimension & Executed realization & Boundary \\
\midrule
Authorization lineage & Every non-root successor has one authorization parent.  Fork atomically partitions one predecessor's residual and fences that predecessor. & No merge transition or multi-parent generation identity is implemented. \\
Manifest commitment & Exactly eight fields are bound: \mcode{codeDigest}, \mcode{controlPath}, \mcode{enforcementConfigDigest}, \mcode{memoryPolicyDigest}, \mcode{modelRouteDigest}, \mcode{policyCompilerDigest}, \mcode{proposerId}, and \mcode{skillClosureDigest}. & A deployment profile must add every further security-relevant surface; an omitted surface is outside this result. \\
Authority algebra & Additive nonnegative integer budgets and opaque, single-holder effect atoms.  The executed budgets are \mcode{actions} and \mcode{tokens}. & No general lattice evaluator, multi-holder effect relation, or arbitrary effect normalizer is implemented. \\
Record binding & Ledger-issued, digest-bound grants, certification records, reservations, events, and receipts are checked against a frozen declared-control-root registry. & The prototype tests record binding and declared-root separation, not caller authentication or key possession; the abstract authenticated-record predicates remain deployment premises. \\
Durability & Intent, temporary state, file-image replacement, and directory synchronization around one handoff. & Atomicity evidence is limited to this file-image replacement protocol; it does not cover distributed consensus or effect-delivery atomicity. \\
\bottomrule
\end{tabularx}
\end{table*}

\subsection{Public source-native refinements}

We next mapped two frozen public systems to the same succession contract while
preserving their native lifecycle evidence.  These are external source
refinements, not additional logical adapters.  In both paths, source identity,
body closure, native validation, and restart observations are measured before
a separate process hosts the prototype ledger and protected-effect broker.

\paragraph{OurArk Genesis/Enoch.}
The run freezes Genesis v0.1.1 at commit
\mcode{e7bb896411db}\allowbreak\mcode{0d82710fab50}\allowbreak
\mcode{c1978b825297844f} and Enoch v0.3.1 at
\mcode{1021e1dacce8}\allowbreak\mcode{5f4a2edebd86}\allowbreak
\mcode{5673671bb37a2142}, both under Apache-2.0,
and exercises their public descent path~\cite{ourark2026}.  The official gate
creates an ephemeral descendant; a second, retained descendant binds the same
parent-at-birth and is used for restart and handoff.  Its manifest maps 176
inherited files: 162 changed content, 14 retained identical content, and no
mapping miss.  A fresh Python process reopened that descendant and discovered
753 inherited tests, with 5 skips and no failure or error.  The broker then
committed exactly one succession, allowed a successor effect, and rejected the
same predecessor operation with \mcode{E\_GENERATION\_INACTIVE}.

\paragraph{Darwin G\"odel Machine.}
The DGM run freezes public commit
\mcode{a565fd2d1dca}\allowbreak\mcode{504ef5104a7c}\allowbreak
\mcode{c0f3bdc4ab9b4fd2} under Apache-2.0 and
measures a 35-file source closure~\cite{dgm2025}.  It invokes the upstream
\mcode{tools/edit.py} EditTool for one deterministic, adapter-authored body
mutation, then launches the copied descendant in a separate Python process.
The upstream editor suite passes 10/10 on both parent and descendant.  A
broader editor/bash compatibility run is 21/22 on both bodies; the same
interactive-shell startup diagnostic causes the single failure and is not a
certification predicate.  The prototype broker again activates the successor
and rejects the predecessor with \mcode{E\_GENERATION\_INACTIVE}.

The full open-ended DGM evolution loop was not executed: it requires external
model-service credentials, container execution, network access, and downloaded
benchmark assets excluded from this offline profile.  The observed DGM event
is therefore one real upstream EditTool mutation plus one independent restart,
not an autonomous open-ended evolution.

Each external adapter replays the 32 registered cases against its one measured
retained descendant or descendant seal.  Both obtain 32/32, with 14 allows and
18 denies each.  The combined external count is thus 64 source-bound profile
replays (28 allows and 36 denies), not 64 independent descents, mutations, or
evolutions.  Neither result instruments the upstream runtime.  The broker is
an out-of-process authorization-succession prototype, and the observations establish
neither an upstream integration, a production deployment, nor endorsement by
OurArk or the DGM authors.

\subsection{Independent seals and corpus boundary}

Table~\ref{tab:evidence-seals} records the three aggregate report digests.  A
dedicated verifier recomputes each bundle's internal report chain and fixed
denominators.  The two external verifiers additionally reject local absolute
paths in generated reports; the DGM verifier also rejects credential- and
secret-like material.

\begin{table*}[t]
\caption{Independent aggregate seals for the frozen evidence bundles.}
\label{tab:evidence-seals}
\small
\begin{tabularx}{\textwidth}{L{0.20\textwidth}Y}
\toprule
Evidence bundle & Domain-separated aggregate report digest \\
\midrule
Provider-free core & \url{sha256:d92b3eefe1a7de48e936505c828ceff3bf83ed831a33e3bedef16cd4c3cb98ef} \\
OurArk Genesis/Enoch & \url{sha256:02e26bac1a4e05855f9be4ed750a9e38de84236f6d18c19f9d0888730e4e64f8} \\
Darwin G\"odel Machine & \url{sha256:21b26c421d36d827b38bc89cfa0df301b477ee45da4f084468de25f002c73851} \\
\bottomrule
\end{tabularx}
\end{table*}

Finally, the external tampering source identified in the evaluation protocol
provides a useful edit-role taxonomy~\cite{harnesstampering2026}, but no
official, licensed machine-readable corpus was frozen for this evaluation.
Accordingly, it contributes zero corpus records and no empirical denominator.
The 28 semantic mutants above remain registered core regressions and are not
relabeled as externally sourced tampering samples.

\section{Related Work}
\label{sec:related}

\subsection{Self-modification, versioned bodies, and guarded synthesis}

Darwin G\"odel Machine demonstrates open-ended evolution in which coding agents
modify their own code and successful variants remain in an archive
\cite{dgm2025}. OurArk treats code, prompts, tools, skills, policies, tests, and
evolution mechanisms as an identity-bearing, versioned body; its Genesis and
Enoch artifacts demonstrate governed changes and recursive descent
\cite{ourark2026}. SEVerA formulates self-evolving agent synthesis with formal
output contracts and verified fallbacks, preserving specified behavioral
constraints during optimization \cite{severa2026}. RegenHarness adds an
evidence-gated recursive-improvement loop for a robot harness: candidate
configuration changes pass fixed regressions and release authorization before
versioned rollout or rollback, and the system is exercised on a real quadruped
\cite{regenharness2026}. These systems supply
concrete mutation, selection, identity, descent, and verification substrates.
Our protocol governs the distinct authorization transition among concurrently
live generations produced by such substrates.

Security analyses show why this boundary matters. The MLAS framework maps
self-evolution threats across functional modules and lifecycle stages and
documents lineage-persistent amplification \cite{selfevolvingsafety2026}.
Auditing Harness Tampering develops a taxonomy and annotated edit pairs for
changes that compromise authorization, provenance, completeness, or evaluation
integrity \cite{harnesstampering2026}. Bad Genius instead challenges candidate
harness improvements with valid counterfactual benchmark transformations and
retains confirmed transformations in a finite archive, reducing dependence on
task-specific shortcuts \cite{badgenius2026}. We use these works to define mutation
surfaces and evidence obligations. Detection of malicious code or proof of
program semantics remains an input to, rather than a substitute for,
population authorization. Neither candidate-validation method allocates
authority across concurrently live generations or linearizes their succession.

\subsection{Mutation-aware authorization continuity}

Earned Authority directly addresses whether an evolving agent may retain a
grant after a transition. It binds authority to a grant-fixed transition
envelope and immutable effect ceiling, covering multiple classes of security-
relevant change and proving conditional non-amplification
\cite{earnedauthority2026}. That is the strongest direct predecessor for
single-generation continuity. Its stated boundary leaves cross-generation
revocation and population-level authority open. Our protocol begins after a
candidate's transition evidence is available and governs sibling allocation,
generation handoff, lineage cuts, independent re-root, and rollback. A
transition-envelope certificate can occupy one typed evidence slot in an
authorization transfer.

Self-Evolving Agents with Anytime-Valid Certificates controls update admission
against a fixed statistical error budget \cite{sea2026}. Like formal
transition verification, that certificate is a strong candidate-evidence
input. The succession question begins when one or more admitted candidates
would receive, share, replace, or lose protected-effect authority.

Exact-hash authorization lies at the conservative endpoint of this design
space: any byte change forces fresh authorization. It prevents grant reuse but
rejects a safe narrowing or semantics-preserving successor even after the old
generation is durably fenced. Mutation-aware continuity preserves more useful
successors, while our protocol adds the population transaction required when
more than one generation can exist.

Governing Dynamic Capabilities binds an agent certificate to a commitment over
its model, tools, and permissions, invalidates access after a capability-set
change, and records signed hash-linked interactions in a trust tree
\cite{dynamiccapabilities2026}. These are direct precedents for capability-bound
identity and change invalidation. This work addresses the next population
state: residual authority is allocated without copying, the predecessor is
fenced at the successor linearization point, ancestor cuts govern dependent
generations, and rollback cannot remint a retired generation grant.

\subsection{Recursive-agent risk budgets}

Progressive Risk Vesting (PRV) holds a trajectory-level risk budget in escrow
and charges recursively spawned branches when they cross an irreversible
activation boundary, proving an anytime probabilistic harm bound under
history-conditional local certificates \cite{prv2026}. It establishes that
branching authority must be controlled at the trajectory level rather than by
independent local risk estimates. This work addresses a complementary
deterministic authorization object: complete effect relations, consumable
resources, concurrency shares, generation identity, durable predecessor
fencing, ancestor cuts, independent re-root, and rollback non-reminting. PRV's
risk charge can be represented as one quantitative component of a population
envelope; it does not replace the generation-lifetime and revocation state used
here.

Agent Contracts gives a formal multidimensional resource-contract framework
with conservation across hierarchical, parallel, and recursive delegation;
retrieval-conditioned topology selection proves a six-dimensional budget
algebra for changing multi-agent topologies; and Token Budgets uses affine
ownership to prevent clone, double-spend, and use-after-delegation failures
\cite{agentcontracts2026,topologybudget2026,tokenbudgets2026}. These works occupy
budget conservation as a standalone contribution. Our resource vectors reuse
that principle inside a generation-succession state whose additional objects
are complete effect relations, predecessor fencing, lineage cuts, independent
re-root, and rollback lifetime.

\subsection{Delegation, ancestry, and hierarchical revocation}

Authorization propagation systems carry user and policy constraints through
multi-agent workflows \cite{authprop2026}. Bounded-agent designs attenuate
delegation and constrain composition \cite{boundedagents2026}. VERA records
federated delegation edges and supplies verifiable revocation across
organizational workflow boundaries \cite{vera2026}. Attested delegation
ancestry binds outputs to signed delegation paths, strengthening provenance
for downstream verification \cite{attestingdelegation2026,lineageuma2026}.
ResidualAuth further characterizes why future-correct revocation requires
authorization history beyond current reachability \cite{residualauth2026}.
These mechanisms
provide compatible authority and evidence inputs. Their principal object is a
delegation or workflow edge rather than a self-modifying generation lifetime
whose predecessor, siblings, rollback source, and body identity change under
one population allocation.

Heartbeat-Bound Hierarchical Credentials (HBHC) already establishes cascading
revocation in an agent hierarchy by requiring periodic parent liveness proofs
and bounding the resulting zombie window under its timing and key assumptions
\cite{hbhc2026}. Accordingly, ancestor revocation by itself is not claimed as a
novel concept here. The contribution is its integration with deterministic
generation identity, aggregate population allocation, atomic replacement,
independent re-root, and rollback replay protection. HBHC can serve as an
efficient credential realization of a lineage cut when its heartbeat and
clock premises match the succession profile.

Authenticated Delegation and Authorized AI Agents separates agent and user
identity in delegated action chains \cite{authenticateddelegation2025}. This
supports the paper's distinction among a generation principal, its operator,
and the authority root from which its allocation derives.

\subsection{Effect transactions, cancellation, and rollback}

Stop Means Stop measures sibling leaks, cancellation orphans, timeout zombies,
and replay, and supplies an external fence-on-cancel control
\cite{stopmeansstop2026}. Cordon provides task-scoped semantic transactions for
staging and validating composed tool effects \cite{cordon2026}; AID-Guard
provides a stateful request-to-provider lifecycle with reservation and retry
control \cite{aidguard2026}. Effect Closure characterizes when a prior
authorization has lost its remaining provider-side path to an effect
\cite{effectclosure2026}. Root-scoped authorization quiescence additionally
accounts for every manifested pre-cut carrier under a retired root epoch and
composes provider-frontier certificates while preserving independently
sufficient support \cite{zhu2026quiescence}. These systems govern run cancellation, tool-effect
transactions, or provider settlement. Our atomic transaction instead replaces
the set of grant-holding software generations; the lower-level controls can
serve as its effect and retirement leaves. In particular, root-scoped
quiescence can retire the old authorization carriers around a cut, whereas the
succession protocol decides how authority is allocated among predecessor,
successor, sibling, rollback, and independently re-rooted generations.

ACRFence directly studies action replay and authority resurrection after agent
checkpoint restore, while Agent libOS combines capability-controlled
processes, parent--child lineage, ceilings, budgets, checkpoint images, restore
intersection, and prepare--dispatch--settle execution
\cite{acrfence2026,agentlibos2026}. Autogenesis likewise provides an auditable
versioned evolution and rollback lifecycle \cite{autogenesis2026}. These works
occupy general checkpoint/rollback defense and several runtime primitives. The
residual result here is a generation-population invariant: equal historical
bytes acquire a fresh generation identity and cannot remint the retired
allocation while sibling and ancestor state are atomically conserved.

\subsection{External enforcement and cross-substrate authority}

An unfireable safety kernel places protected effects behind an out-of-process
path that mutable agents cannot remove \cite{unfireable2026}. Cross-substrate
authority work likewise argues that authority state must survive movement
among agent runtimes and orchestration surfaces \cite{crosssubstrate2026}.
Our protocol assumes such complete mediation and specifies the generation
and population state that the external path checks. The kernel supplies
placement and availability of the control point; it does not decide how a
parent allocation is divided, how a replacement is linearized, or why a
historical body lacks its historical grant.

Work on epistemic Sybil resistance tracks evidence ancestry to prevent copied
or correlated evidence from appearing independent \cite{epistemicsybil2026}.
Our evidence-root predicate adopts the same general lesson at an authorization
boundary: principal counts are insufficient when declared principals share one control root.
The result here is limited to authenticated declared control provenance and
does not infer hidden organizational or economic coordination.

\subsection{Capability safety and revocation}

Capability systems established unforgeable authority tokens, delegation by
possession, attenuation, and the importance of controlling the complete path
to an object \cite{dennis1966,saltzer1975,abadi1993}. Contextual caveats support
decentralized attenuation \cite{macaroons2014}, and leases provide the classic
time-bounded validity pattern \cite{gray1989}. Revocation research adds
indirection, versioning, expiry, and online validation to invalidate delegated
authority \cite{hagstrom2001}. Our protocol uses
these principles but makes the revocation and transfer subject a single-parent
generation forest with a changing active population. Its distinctive state is the joint
allocation and atomic succession record: a valid token from one generation is
insufficient unless the generation, current authorization root, lineage cut
set, population version, and effect allocation are current at commitment.

\subsection{Boundary summary}

Table~\ref{tab:related-boundary} summarizes the nearest mechanisms. The rows
are composable rather than mutually exclusive; the final column identifies the
state added by this paper.

\begin{table*}[t]
\caption{Boundary to the closest lines of work.}
\label{tab:related-boundary}
\footnotesize
\begin{tabularx}{\textwidth}{L{0.19\textwidth}L{0.25\textwidth}L{0.25\textwidth}Y}
\toprule
Line of work & Principal object & Established control & State used by this work \\
\midrule
Earned Authority & one evolving grant holder & transition envelope and fixed effect ceiling & active generation family and allocation transaction \\
PRV & recursive activation trajectory & probabilistic risk escrow and anytime harm bound & deterministic typed effects, resources, epochs, and lifetime \\
HBHC / VERA & hierarchical credentials / delegation edges & cascading or verifiable revocation & generation cut joined with handoff, re-root, and rollback \\
OurArk / DGM & versioned descent / mutation archive & identity-bearing evolution and candidate selection & protected-effect authority assigned to current generations \\
SEVerA / RegenHarness / Bad Genius & synthesized program / versioned harness candidate & constrained outputs, evidence-gated rollout, and counterfactual validation & external succession of executable generations \\
Root-scoped quiescence & retired authorization-root carriers & pre-cut accounting and post-fence exclusion & allocation and replacement of generation principals \\
External safety kernel & protected execution path & non-removable mediation & population and lineage predicates evaluated on that path \\
\bottomrule
\end{tabularx}
\end{table*}

\section{Assurance Boundary and Artifact Scope}
\label{sec:assurance}

\subsection{What the result establishes}

The formal result establishes a conditional safety refinement from the
succession protocol to mediated protected effects. Under assumptions
A1--A10, every accepted effect has a current active generation and authority
witness; aggregate active descendant authority remains below its root grant
and the persistent population ceiling; a cut ancestor cannot support a later accepted effect; replacement
has no committed dual-active state; historical artifacts do not reactivate
historical authority; and candidate-controlled evidence cannot fill required
independent slots.

The executable result has a finite, named denominator. It establishes that the
reference ledger and two logical mappings realize the registered transitions
and that a separately implemented checker detects the registered semantic
violations after the enclosing integrity records are consistently resealed.
These are
mechanism and refinement results over the recorded corpus and fault space.
They are reported as exact counts rather than population estimates.
Both the core model and executable result use the frozen single-authorization-
parent profile. A fork may activate several children, each with that same
predecessor as its sole authorization parent, while atomically fencing the
predecessor. Multi-parent authority merge is outside the formal, implemented,
and executed result.

\subsection{Inputs governed outside the theorem}

Several inputs have their own assurance process. The effect normalizer must
retain every policy-relevant field and correlation. The generation manifest
must cover every security-relevant body and runtime surface. Evidence roots
must be admitted under accurate control provenance. The broker manifest must
cover every protected commitment path. A substantive code verifier or human
reviewer must correctly evaluate the predicates it evaluates and records. The protocol
authenticates, composes, conserves, and revokes those decisions; it does not
replace their domain semantics.

Physical consequences already committed before a valid cut remain historical
effects. Compensation, legal settlement, task success, model alignment, and
general semantic equivalence are separate properties. The protocol prevents
new mediated authority from being derived from a retired lineage; it does not
rewrite the past.

\subsection{Failure classification}

The protocol semantics distinguishes three outcomes rather than permitting an
optimistic fallback:

\begin{itemize}
  \item \Deny{} applies when current evidence proves a policy violation, stale
  lineage, insufficient allocation, cut ancestor, stale lifecycle/transfer
  replay, effect-request conflict, or invalid transition;
  \item \Indeterminate{} applies when a required identity, lineage edge,
  current ledger fact, normalizer field, evidence root, or broker observation
  is unavailable or conflicting; and
  \item \Allow{} is available only after all registered predicates hold in the
  same serialized state used to reserve or commit the protected effect.
\end{itemize}

These outcomes classify fresh authorization decisions. Under A10, an exact
retry of an already committed effect request returns the prior success receipt
without a new admission or debit; it is neither a fresh \Allow{} nor a denied
request.

The bounded reference implementation represents every missing, conflicting, or
unsupported fact as a typed rejection code; it does not expose a separate
\Indeterminate{} transport state or a reservation-expiry scheduler. A deployment
that adds timeout or operator routing must preserve the same rule: absence of
evidence cannot activate a candidate, and every fallback route needs its own
explicit allocation.

\subsection{Evidence classes}

Table~\ref{tab:evidence-classes} keeps proof, execution, and external
refinement distinct.

\begin{table*}[t]
\caption{Evidence classes and their admissible interpretation.}
\label{tab:evidence-classes}
\small
\begin{tabularx}{\textwidth}{L{0.20\textwidth}L{0.30\textwidth}Y}
\toprule
Evidence class & Object & Supported statement \\
\midrule
Formal & definitions, assumptions, lemmas, theorems, proofs & conditional protocol safety for all executions satisfying the model premises \\
Reference execution & registered finite corpus, ledger, brokers, two logical adapters & implementation conformance over the executed mutation and lineage cases \\
Adversarial verification & independent checker and consistently resealed semantic regressions & checker separation and detection of the registered trace-level violations \\
Fault execution & durable crash cuts and deterministic schedules & recovery and atomicity over the exact enumerated interruption/interleaving space \\
External refinement & frozen public runtime commits, native descent or edit, fresh restart, and broker traces & field mapping and boundary behavior for the specifically executed versions \\
\bottomrule
\end{tabularx}
\end{table*}

No evidence class is relabeled as another. In particular, logical adapter names
describe runtime shapes and never imply execution of a named public project.
External results enter the paper only with immutable source versions, commands,
mapping commitments, and fresh traces.

\subsection{Artifact publication boundary}

The research abstraction contains generation identities, typed effects,
allocations, root epochs, generation sequences, cuts, and receipts. It requires no customer data, tenant
configuration, production topology, credentials, proprietary policy corpus,
or company-internal deployment detail. Evaluation records use inert synthetic
effect tuples and public-source metadata where applicable.

The public preprint source archive is generated by a manuscript-only whitelist. It
contains the reviewed \LaTeX{} sources, bibliography, and only assets required
to compile the paper. Executable artifacts, adapters, checker traces, temporary
files, source caches, and non-manuscript research materials remain outside that
archive. Public-source attribution identifies each upstream repository and
immutable version without embedding local workspace paths.

\section{Conclusion}
\label{sec:conclusion}

Self-modification changes more than one agent's code. Forking, promotion,
coordination among retained branches, delayed descendants, and rollback change
the set of principals that can exercise a shared authorization. Per-generation
acceptance therefore cannot by itself preserve a population-wide bound.

The proposed protocol makes that bound explicit. Canonical hash-derived generation
principals separate body identity from authorization lifetime; an external
population ledger allocates rather than copies effect relations and resource
shares; atomic transactions replace or rebalance the active frontier; lineage
cuts reach every dependent descendant; independent re-root preserves genuinely
new authority; and rollback creates a fresh generation instead of reviving a
historical grant. Protected effects revalidate this state at their commitment
point.

The formal results connect these mechanisms to population-safe succession,
fork conservation, revocation closure, atomic handoff, rollback non-reminting,
and evidence-root independence. The registered executable evaluation makes the
separating executions concrete across two logical runtime mappings, durable
fault cuts, interleavings, and independently checked semantic mutations.
Separate OurArk and Darwin G\"odel Machine adapters further exercise frozen
public source-native descent or edit, fresh-process restart, atomic handoff,
predecessor rejection, and all 32 registered decisions without treating replay
count as evolution count. Together, the protocol and evidence establish
authorization succession as a first-class control plane for evolving agent
populations.

\appendix
\section{Proofs of the Succession Results}
\label{app:proofs}

We reason over finite prefixes of the accepted transition relation in
Section~\ref{sec:formal}. A rejected command may append a diagnostic event,
but it changes no generation state, cut set, allocation, reservation,
authorization-object consumption record, or protected-effect commitment. The
cryptographic and systems facts used below are exactly A1--A10.

\subsection{Trace notation and elementary facts}

For the remainder of the appendix, every trace starts in a state satisfying
\(\mathsf{InitOK}(S_0)\) from Definition~\ref{def:initial-state}. Write
\[
 S_0\xrightarrow{o_1}S_1\xrightarrow{o_2}\cdots
 \xrightarrow{o_n}S_n
\]
for an accepted trace. By A6, every security-changing transition has one
position in the external lineage ledger's serial order, written \(<_{L}\). An
effect admitted at such a position may settle later; its authorization remains
attributable to that admission. A10 applies only to a protected-effect request:
an exact retry returns its existing receipt and creates no second admission,
effect, or debit. No proof below assumes exact-result retry for another command
class.

The operation \(\oplus\) of \eqref{eq:authority-algebra} is associative and commutative: qualitative
components use the finite join, while budget and exposure components use
vector addition. It is monotone in each argument under \(\sqsubseteq\). For
every registered population profile, A5 requires \(\Phi_p\) to be downward closed
under removal of a labeled allocation and attenuation of an allocation.
Removing an active residual leaves \(c_r(S)\) unchanged and can only decrease
the aggregated components of both \eqref{eq:root-ok} and
\eqref{eq:population-ok}. A fresh debit \(\delta\) increases \(c_r(S)\) and
decreases the admitting generation's \(b_g(S)\) by the same \(\delta\), so the
root-lifetime sum is invariant while current population exposure decreases.
These facts preserve \(\mathsf{Safe}_r\).

\begin{lemma}[Canonical generation separation]
\label{lem:generation-separation}
Within a population, two accepted generation registrations never have the
same population-issued sequence number. If their population, authorization
root, root epoch, root-grant digest, unique parent, complete lineage-atom set,
manifest digest, or sequence differs, their generation identifiers differ
except with a digest collision. In particular, a rollback and an independent
re-root have identifiers distinct from the corresponding historical
generation.
\end{lemma}

\begin{proof}
A3 makes allocation of \(n_g\) monotone and durable, and the ledger's
registration guard rejects reuse. Equation~\eqref{eq:generation-identity} canonically encodes the domain
separator, population, authorization root, root epoch, root-grant digest,
sequence, unique authorization parent, complete lineage-atom set, and manifest
digest as distinct typed fields. Distinct inputs have distinct encodings;
equality of their digests would be a collision excluded by A2.
Definition~\ref{def:rollback} requires a fresh sequence and current parent, so
the rollback and historical inputs differ even when their manifests are equal.
Definition~\ref{def:reroot} additionally gives a fresh authorization root,
root epoch, grant digest, sequence, and root lineage atom.
\end{proof}

\begin{lemma}[Pre-activation inertness]
\label{lem:preactivation}
Candidate construction, quarantine execution, evidence recording, and a held
singleton reservation do not add a member to \(\mathsf{AD}_S(r)\) and cannot
commit a protected effect. An atomic fork child acquires authority only at its
family commit position.
\end{lemma}

\begin{proof}
The staged path leaves a candidate \(\mathsf{PROPOSED}\) or
\(\mathsf{RESERVED}\); an evidence record does not itself change that state.
The fork path has no active child record before its atomic commit.
Definition~\ref{def:active-descendants} requires \(\mathsf{ACTIVE}\), and A1
independently requires the broker to read an exact current active record before
a fresh admission. A reservation or inline fork description is an
authorization input, not an executable handle, and cannot satisfy that check.
\end{proof}

\begin{lemma}[Handoff preservation]
\label{lem:handoff-preservation}
Suppose I1--I7 hold in \(S\). If
\(\mathsf{CommitFamily}(p,N,\chi)\) is accepted, they hold in its post-state
\(S'\).
\end{lemma}

\begin{proof}
The guard re-resolves each child manifest, assigns exactly one parent \(p\),
and recomputes \eqref{eq:generation-identity}, establishing I1. It checks each independent evidence
record with \eqref{eq:cert-ok} and the family authorization object with
\eqref{eq:auth-ok}. For a singleton, the held reservation binds the current
predecessor residual and population version and is consumed at commit; for a
fork, the complete family bundle is validated inline at the same atomic commit
that fences the predecessor. Thus each child is activated by one accepted
\(\mathsf{AuthOK}\) family commit, establishing I2 without asserting a
persisted, consumed fork bundle.

For every \(g\in N\), the guard constructs
\(A_g=A_p\cup\{a_g\}\) from authoritative records and checks
\(A_g\cap\mathsf{Cut}_S(p(r))=\varnothing\). There is no second parent from which an atom
could be omitted. This establishes I3. The transaction rejects reused
generation identifiers and authorization objects; for the staged path it also
requires an unspent reservation. A6 makes their consumption and all state
writes atomic. No appeal to A10 is needed, so I4 holds.

Before committing, the guard constructs the exact post-state population:
unrelated active members plus \(N\), with \(p\) removed. Acceptance requires
\(\mathsf{Safe}_r(S',M')\) on that labeled allocation family and
\eqref{eq:single-transfer} or \eqref{eq:fork-transfer} against
\(p\)'s reconstructed residual. The handoff creates no effect receipt, so
\(c_r(S')=c_r(S)\). Other roots are unchanged; hence I5 holds for every root.
Existing instances of I6 remain true. At one commit position, the transaction
marks \(p\) \(\mathsf{FENCED}\) and all members of \(N\)
\(\mathsf{ACTIVE}\). It neither mutates \(p\)'s immutable grant nor restores a
terminal record, establishing I7.
\end{proof}

\begin{lemma}[Cuts and effect requests preserve the invariants]
\label{lem:cut-effect-preservation}
If I1--I7 hold before an accepted ancestor cut, fresh protected-effect
admission, or exact effect-request replay, they hold afterward.
\end{lemma}

\begin{proof}
An ancestor cut adds one atom \(a\) to \(\mathsf{Cut}_S(p(r))\) under A3 and A9. It creates no
identity, evidence record, allocation, or effect receipt, so I1, I2, I4, and
I6 are unchanged. Every nonterminal registered generation whose closure
contains \(a\) is marked \(\mathsf{REVOKED}\); already terminal generations
remain terminal, and any unobserved descendant still fails
\eqref{eq:current}
because A4 requires the complete unique-parent closure. Thus I3 and I7 hold.
The current active population can only lose members while every historical
debit remains in \(c_r\), so downward closure preserves I5.

For a fresh effect request, A1 checks one exact current active generation, its
complete atom closure and current cut set, remaining allocation, normalized
effect, and budget at the admission position. A fresh acceptance writes one
receipt and one debit in the same transaction under A6 and records the request
identity under A10. This establishes I6 and preserves I4. For debit
\(\delta\), \(c_r\) increases by \(\delta\) while the active
\(b_g(S)\) decreases by \(\delta\). The budget conjunct of
\eqref{eq:root-ok} is unchanged, the budget conjunct of
\eqref{eq:population-ok} decreases, and every other conjunct is preserved. No
identity, parent edge, state, or cut set is widened.

For an exact retry, A10 returns the previously bound receipt. It appends no
protected-effect commitment and no debit, even when the named generation is no
longer active. Hence all seven invariants are unchanged. Reusing the identity
with a different canonical effect request is rejected and is outside the
accepted relation.
\end{proof}

\begin{lemma}[Independent re-root preservation]
\label{lem:reroot-preservation}
If I1--I7 hold in \(S\) and an independent re-root is accepted, they hold in
its post-state \(S'\).
\end{lemma}

\begin{proof}
Definition~\ref{def:reroot} first requires an empty raw active frontier for the
affected population. The transition therefore transfers no old-root residual
and cannot overlap an old active allocation. It authenticates a fresh external
root grant under A2, assigns a strictly fresh root epoch and generation
sequence under A3, resolves the exact manifest, and creates \(g'\) with no
authorization parent. These checks establish I1. Accepted independent evidence
and the new root-grant authorization establish I2. The constructed closure
contains only the fresh root and generation atoms and is checked against the
cut set, establishing I3.

The root, grant, generation, and lineage identifiers are fresh, and the new
certification record binds accepted evidence to the fresh generation even when
the opaque evidence digest has appeared before. The transaction creates each
new identifier once at one serial position; this establishes I4 without using
A10. Before changing the current-root pointer, the guard checks
the new grant and \(\mathsf{Safe}_{r'}\) on the exact post-state containing
\(g'\) as the population's sole active generation. The new root has zero prior
lifetime debit, old roots expose no active descendants, and other populations
are unchanged, so I5 holds. The transition creates no protected-effect receipt,
leaving I6 unchanged. Finally it atomically advances
\(\mathsf{curRoot}_S(p_0)\), retains every historical root and terminal record,
and never reactivates an old generation or cut atom. Monotonic root epochs and
records under A3 establish I7.
\end{proof}

\begin{lemma}[All remaining accepted transitions preserve the invariants]
\label{lem:other-preservation}
Assuming I1--I7, every accepted transition other than family handoff, ancestor
cut, fresh effect admission, exact effect replay, or independent re-root
preserves them.
\end{lemma}

\begin{proof}
By Lemma~\ref{lem:preactivation}, proposal, quarantine execution, evidence
recording, and reservation add no current active member and no effect receipt.
Their guards bind identities and evidence prospectively, and phase, version,
uniqueness, and monotonicity checks reject stale or duplicate commands. This
argument uses A3 and A6, not a general retry property.

Voluntary termination or an administrative fence removes a member from the
current active population and does not release a consumed authorization
object. Population safety follows by downward closure, while fence monotonicity
follows from A3 and A6. A trusted bootstrap root is admitted only with a
canonical identity, a profile-valid authenticated bootstrap grant whose issuer
control root differs from the proposer control root, fresh allocation, zero
initial lifetime debit, and a post-state satisfying
\(\mathsf{Safe}_r(S',M')\). Its root is installed as the population's unique
current root. These facts establish the bootstrap branch of I2 without
postulating a separate certification record. No other registered transition
changes protected authority.
\end{proof}

\begin{lemma}[Invariant preservation]
\label{lem:all-invariants}
Under A1--A10, I1--I7 hold after every prefix of every accepted trace that
starts from \(\mathsf{InitOK}(S_0)\).
\end{lemma}

\begin{proof}
By Definition~\ref{def:initial-state}, an empty initial state satisfies I1--I7
vacuously. In every provisioned initial population, the canonical unique root
record establishes I1 and I3; its profile-valid authenticated bootstrap grant
with issuer/proposer control-root separation establishes the bootstrap branch
of I2; unique counters and empty consumption state establish I4; the explicit
\(\mathsf{Safe}_r\) guard establishes I5; I6 is vacuous because no effect
receipt exists; and fresh records and an empty cut set establish I7. Disjoint
populations compose these facts. Induct on trace length.
Lemma~\ref{lem:handoff-preservation} handles a family handoff,
Lemma~\ref{lem:cut-effect-preservation} handles a cut, effect admission, or
effect replay, Lemma~\ref{lem:reroot-preservation} handles independent re-root,
and Lemma~\ref{lem:other-preservation} handles every remaining accepted step.
\end{proof}

\subsection{Proofs of the six main results}

\begin{proof}[Proof of Theorem~\ref{thm:population-safe}]
Lemma~\ref{lem:all-invariants} gives I5 in every reachable state. Expanding
\(\mathsf{Safe}_r\) by Definition~\ref{def:population-ceiling} yields both
classes of conjuncts in \eqref{eq:population-bound}. The \(\mathsf{RootOK}\)
conjuncts bound qualitative authority, root-lifetime debit plus active
remaining budget, and holder exposure by \(G_r\). The \(\mathsf{PopOK}\)
conjuncts independently bound the current active allocation family and its
relational predicate by persistent \(C_{p(r)}\). In particular, fencing a
spender cannot remove its debit from \(c_r\), so a successor cannot recover
consumed root budget; an independent re-root cannot raise current population
exposure above its persistent ceiling.

For the attribution statement, I6 gives a digest-bound receipt for one exact
\(\mathsf{gid}(g)\), atom-set digest, source-allocation digest, canonical
operation, normalized effect, budget debit, and ledger position. A1 validates
that \(g\) is active and current at that position and that the effect and debit
lie within its allocation. The request index and A10 prevent a second fresh
commitment under the same effect-request identity; an exact retry only returns
the existing receipt. A2 prevents substitution of a different canonical
object under a recorded digest. Thus each protected commitment has exactly one
current generation witness at admission.
\end{proof}

\begin{proof}[Proof of Theorem~\ref{thm:fork-conservation}]
Fix an accepted fork from \(p\) to \(N\). Equation~\eqref{eq:fork-transfer}
directly gives
\begin{equation}
\label{eq:fork-bound-proof}
 \bigoplus_{g\in N}\widehat\alpha_g
 \sqsubseteq \alpha_p(S).
\end{equation}
The family commit fences \(p\) at the same position that activates \(N\), and
A3 prevents \(p\)'s residual from becoming active again. A6 admits that
predecessor state to only one serialized consuming family transition. The fork
therefore replaces, rather than supplements, the predecessor allocation.

For a fork tree, induct on accepted fork positions among the current active
leaves.
Equation~\eqref{eq:fork-bound-proof} replaces one
current allocation by a bounded child composition at each step; associativity,
monotonicity, and transitivity preserve the originating bound. If two
descendants both receive a counted qualitative effect, the abstract profile
assigns each a positive holder unit, so vector addition accounts for both. In
the reference profile, the stronger pairwise-disjoint effect-label check
rejects that duplicate ownership directly. Numeric child budgets are summed
componentwise in both cases. Thus neither a direct nor nested fork amplifies
the originating residual.
\end{proof}

\begin{proof}[Proof of Theorem~\ref{thm:revocation-closure}]
Let cut \(c\) insert atom \(a\) into \(\mathsf{Cut}_S(p(r))\). By A3 and A9 the insertion is
durable and monotone. Consider a fresh effect admission \(k\) with
\(c<_{L}k\) and a presenting generation \(g\) for which \(a\in A_g\). At
\(k\), A1 and A9 require \(\mathsf{Current}_S(g)\).
Equation~\eqref{eq:current} requires
\(A_g\cap\mathsf{Cut}_S(p(r))=\varnothing\), yet the intersection contains \(a\), a
contradiction.

By A4 and \eqref{eq:generation-identity}, every ordinary descendant contains the transitive unique-parent
closure; inserting an intermediate generation cannot remove an ancestor atom.
Hence the contradiction applies at arbitrary descendant depth.
Definition~\ref{def:reroot} explains the stated exception: an independent
re-root has a new external authority cause and no old authorization-parent
atom. An exact retry after \(c\) may return the receipt of an effect admitted
before \(c\), but A10 creates no new admission, effect, or debit and therefore
does not contradict the result.
\end{proof}

\begin{proof}[Proof of Theorem~\ref{thm:atomic-handoff}]
Before \(\lambda_h\), every staged child is non-active and every inline fork
child lacks an active record; Lemma~\ref{lem:preactivation} and A1 exclude
their production effects. At \(\lambda_h\),
Definition~\ref{def:handoff} fences \(p\) and activates all members of \(N\) in
one durable transaction. There is no observable accepted state between these
writes. After \(\lambda_h\), I7 prevents \(p\)'s reactivation, and A1 rejects
its fresh effect admission.

A6 serializes every fresh protected-effect admission with the family handoff.
An admission therefore lies either before or after \(\lambda_h\), never both
and never between its component writes. The earlier side can contain
exercisable predecessor authority but no child authority; the later side can
contain child authority but no predecessor authority. A pre-handoff commitment
that settles later remains attributed to its earlier admission. An exact retry
may return that old receipt after the handoff, but it is non-consuming and is
not a second post-handoff admission under A10.
\end{proof}

\begin{proof}[Proof of Theorem~\ref{thm:rollback}]
Lemma~\ref{lem:generation-separation} gives
\(\mathsf{gid}(g_{rb})\ne\mathsf{gid}(a)\) even when
\(m_{g_{rb}}=m_a\). Equation~\eqref{eq:cert-ok} binds an evidence record to the exact
generation identifier and manifest digest. Equation~\eqref{eq:auth-ok} binds the family
authorization object to the current predecessor, child identifier, evidence,
allocation, and residual state. An old evidence record, reservation, or
authorization object therefore fails the rollback's exact binding even when
the manifest bytes match.

I4 prevents reuse of a consumed reservation or authorization identifier, while
A3 prevents restoring the external ledger to an earlier state. If an ancestor
atom has been cut, Theorem~\ref{thm:revocation-closure} blocks the ordinary
rollback lineage. The only admission path for \(g_{rb}\) is consequently a
fresh ordinary handoff whose unique-parent closure is current.
Lemma~\ref{lem:handoff-preservation} checks its post-state
\(\mathsf{Safe}\) condition, including irreversible lifetime debit and the
persistent population ceiling, so
restored bytes never remint historical authority. If the same historical
manifest is instead restored under an independent re-root, the resulting root
generation is \(g'\), not \(g_{rb}\), and
Lemma~\ref{lem:reroot-preservation} together with
Corollary~\ref{cor:reroot} supplies the corresponding non-laundering result.
\end{proof}

\begin{proof}[Proof of Theorem~\ref{thm:no-self-certification}]
There are two activation paths for a non-bootstrap candidate. Under an
existing root, \(\mathsf{CommitFamily}\) requires
\(\mathsf{CertOK}_S(g,\zeta_g)\) for each child and
\(\mathsf{AuthOK}_S(p,N,\chi)\) for the transition. By
\eqref{eq:cert-ok} and A7, at least one
required certifier control root differs from the proposer control root, and the
candidate cannot modify the frozen registry or remove that slot. By
\eqref{eq:auth-ok} and
the trusted-boundary premise, the lineage authorization decision also
originates outside the candidate boundary.

For independent re-root, Definition~\ref{def:reroot} instead requires both a
fresh authenticated external root grant and accepted independent evidence,
with the registered issuer and evidence control roots outside the proposer
control root. The candidate cannot mint the fresh grant, satisfy the
independent evidence predicate, or change the frozen control-root registry.
A2 excludes producing an accepted external authenticator or substituting a
candidate-controlled principal under any of these registered roots.
Consequently, records originating only in or controlled by \(g\) cannot
satisfy either activation guard. The result establishes
provenance independence of authorization evidence; the factual soundness of
the external observation remains a separate domain-semantic obligation stated
in Section~\ref{sec:assurance}, not a consequence of A7.
\end{proof}

\begin{proof}[Proof of Corollary~\ref{cor:reroot}]
Definition~\ref{def:reroot} gives \(\mathsf{par}(g')=\bot\) under the new root
and records reused artifacts only as provenance. A4 prevents an omitted
authorization parent, while the fresh root grant, root epoch, sequence,
allocation, evidence record, and lineage set contain no old authority atom.
A7 validates the independent evidence root, A8 binds the fresh allocation, and
A9 rejects cut atoms. Lemma~\ref{lem:reroot-preservation} establishes the
unique current-root switch and exact new-root post-state. The Population-Safe
Succession result then attributes an effect of \(g'\) to that exact fresh root
and allocation. A field deletion or label edit supplies neither the
authenticated external root grant nor the new canonical bindings, so A2
rejects it rather than accepting it as a re-root.
\end{proof}

\subsection{Evidence status}

The arguments above are conditional mathematical proofs over the registered
transition system. The executable evidence instantiates them only for the
fixed eight-field manifest, trusted control-root registry and digest records,
single-parent succession, atomic single-predecessor fork, cut set, disjoint
effect labels, and integer budgets. Per-atom epoch vectors, general profile
digests, public-key signatures, arbitrary authority relations, and
multi-parent merge are not inferred from those runs. Executable cases,
independent trace checking, consistently resealed semantic regressions,
logical-adapter replay, crash cuts, contender orders, and public-source runtime
mapping remain distinct evidence obligations reported from frozen result
records.

\bibliographystyle{ACM-Reference-Format}
\setlength{\bibsep}{-0.2pt}
\bibliography{references}


\begin{thebibliography}{42}

%%% the Web address.

\ifx \showCODEN    \undefined \def \showCODEN     #1{\unskip}     \fi
\ifx \showISBNx    \undefined \def \showISBNx     #1{\unskip}     \fi
\ifx \showISBNxiii \undefined \def \showISBNxiii  #1{\unskip}     \fi
\ifx \showISSN     \undefined \def \showISSN      #1{\unskip}     \fi
\ifx \showLCCN     \undefined \def \showLCCN      #1{\unskip}     \fi
\ifx \shownote     \undefined \def \shownote      #1{#1}          \fi
\ifx \showarticletitle \undefined \def \showarticletitle #1{#1}   \fi
\ifx \showURL      \undefined \def \showURL       {\relax}        \fi
% invisible to TeX
\providecommand\bibfield[2]{#2}
\providecommand\bibinfo[2]{#2}
\providecommand\natexlab[1]{#1}
\providecommand\showeprint[2][]{arXiv:#2}

\bibitem[Abadi et~al\mbox{.}(1993)]%
        {abadi1993}
\bibfield{author}{\bibinfo{person}{Mart{\'i}n Abadi}, \bibinfo{person}{Michael
  Burrows}, \bibinfo{person}{Butler Lampson}, {and} \bibinfo{person}{Gordon
  Plotkin}.} \bibinfo{year}{1993}\natexlab{}.
\newblock \showarticletitle{A Calculus for Access Control in Distributed
  Systems}.
\newblock \bibinfo{journal}{\emph{ACM Transactions on Programming Languages and
  Systems}} \bibinfo{volume}{15}, \bibinfo{number}{4} (\bibinfo{year}{1993}),
  \bibinfo{pages}{706--734}.
\newblock
\href{https://doi.org/10.1145/155183.155225}{doi:\nolinkurl{10.1145/155183.155225}}


\bibitem[Banerjee et~al\mbox{.}(2026)]%
        {severa2026}
\bibfield{author}{\bibinfo{person}{Debangshu Banerjee},
  \bibinfo{person}{Changming Xu}, \bibinfo{person}{Eugene Ie},
  \bibinfo{person}{Ming Zhang}, \bibinfo{person}{Daiyi Peng},
  \bibinfo{person}{Chu-Cheng Lin}, {and} \bibinfo{person}{Gagandeep Singh}.}
  \bibinfo{year}{2026}\natexlab{}.
\newblock \bibinfo{title}{{SEVerA}: Verified Synthesis of Self-Evolving
  Agents}.
\newblock
\showeprint[arxiv]{2603.25111}~[cs.LG]
\urldef\tempurl%
\url{https://arxiv.org/abs/2603.25111}
\showURL{%
\tempurl}
\newblock
\shownote{Version 2, revised 24 April 2026}.


\bibitem[Bara(2026)]%
        {epistemicsybil2026}
\bibfield{author}{\bibinfo{person}{Marc Bara}.}
  \bibinfo{year}{2026}\natexlab{}.
\newblock \bibinfo{title}{Epistemic {Sybil} Resistance: Multiplying {AI} Agents
  Without Multiplying Evidence}.
\newblock
\showeprint[arxiv]{2609.01873}~[cs.AI]
\urldef\tempurl%
\url{https://arxiv.org/abs/2609.01873}
\showURL{%
\tempurl}
\newblock
\shownote{Version 1, submitted 1 September 2026}.


\bibitem[Birgisson et~al\mbox{.}(2014)]%
        {macaroons2014}
\bibfield{author}{\bibinfo{person}{Arnar Birgisson}, \bibinfo{person}{Joe~Gibbs
  Politz}, \bibinfo{person}{{\'U}lfar Erlingsson}, \bibinfo{person}{Ankur
  Taly}, \bibinfo{person}{Michael Vrable}, {and} \bibinfo{person}{Mark
  Lentczner}.} \bibinfo{year}{2014}\natexlab{}.
\newblock \showarticletitle{Macaroons: Cookies with Contextual Caveats for
  Decentralized Authorization in the Cloud}. In
  \bibinfo{booktitle}{\emph{Proceedings of the Network and Distributed System
  Security Symposium}}. \bibinfo{publisher}{Internet Society},
  \bibinfo{address}{San Diego, CA, USA}, \bibinfo{numpages}{16}~pages.
\newblock
\showISBNx{1-891562-35-5}
\href{https://doi.org/10.14722/ndss.2014.23212}{doi:\nolinkurl{10.14722/ndss.2014.23212}}


\bibitem[Chen et~al\mbox{.}(2026)]%
        {cordon2026}
\bibfield{author}{\bibinfo{person}{Zheng Chen}, \bibinfo{person}{Hanqing Liu},
  \bibinfo{person}{Duling Xu}, \bibinfo{person}{Dong Dong},
  \bibinfo{person}{Jialin Li}, \bibinfo{person}{Bangzheng Pu}, {and}
  \bibinfo{person}{Jidong Zhai}.} \bibinfo{year}{2026}\natexlab{}.
\newblock \bibinfo{title}{Cordon: Semantic Transactions for Tool-Using {LLM}
  Agents}.
\newblock
\showeprint[arxiv]{2606.17573}~[cs.OS]
\urldef\tempurl%
\url{https://arxiv.org/abs/2606.17573}
\showURL{%
\tempurl}
\newblock
\shownote{Version 1, submitted 16 June 2026}.


\bibitem[Choi et~al\mbox{.}(2026)]%
        {residualauth2026}
\bibfield{author}{\bibinfo{person}{Moonwon Choi}, \bibinfo{person}{Seokho
  Jeong}, {and} \bibinfo{person}{Seunggeun Lee}.}
  \bibinfo{year}{2026}\natexlab{}.
\newblock \bibinfo{title}{{ResidualAuth}: What Authorization State Must
  Language Agents Preserve under Revocable Delegation?}
\newblock
\showeprint[arxiv]{2609.08062}~[cs.AI]
\urldef\tempurl%
\url{https://arxiv.org/abs/2609.08062}
\showURL{%
\tempurl}
\newblock
\shownote{Version 1, submitted 8 September 2026}.


\bibitem[Dennis and Horn(1966)]%
        {dennis1966}
\bibfield{author}{\bibinfo{person}{Jack~B. Dennis} {and} \bibinfo{person}{Earl
  C.~Van Horn}.} \bibinfo{year}{1966}\natexlab{}.
\newblock \showarticletitle{Programming Semantics for Multiprogrammed
  Computations}.
\newblock \bibinfo{journal}{\emph{Commun. ACM}} \bibinfo{volume}{9},
  \bibinfo{number}{3} (\bibinfo{year}{1966}), \bibinfo{pages}{143--155}.
\newblock
\href{https://doi.org/10.1145/365230.365252}{doi:\nolinkurl{10.1145/365230.365252}}


\bibitem[Deochake(2026)]%
        {hbhc2026}
\bibfield{author}{\bibinfo{person}{Saurabh Deochake}.}
  \bibinfo{year}{2026}\natexlab{}.
\newblock \bibinfo{title}{Heartbeat-Bound Hierarchical Credentials:
  Cryptographic Revocation for {AI} Agent Swarms}.
\newblock
\showeprint[arxiv]{2605.20704}~[cs.CR]
\urldef\tempurl%
\url{https://arxiv.org/abs/2605.20704}
\showURL{%
\tempurl}
\newblock
\shownote{Version 1, submitted 20 May 2026}.


\bibitem[Dobrin and {{\L}ukasz Chmiel}(2026)]%
        {unfireable2026}
\bibfield{author}{\bibinfo{person}{Seth Dobrin} {and}
  \bibinfo{person}{{{\L}ukasz Chmiel}}.} \bibinfo{year}{2026}\natexlab{}.
\newblock \bibinfo{title}{The Unfireable Safety Kernel: Execution-Time {AI}
  Alignment for {AI} Agents and Other Escapable {AI} Systems}.
\newblock
\showeprint[arxiv]{2606.26057}~[cs.AI]
\urldef\tempurl%
\url{https://arxiv.org/abs/2606.26057}
\showURL{%
\tempurl}
\newblock
\shownote{Version 1, submitted 24 June 2026}.


\bibitem[Gray and Cheriton(1989)]%
        {gray1989}
\bibfield{author}{\bibinfo{person}{Cary~G. Gray} {and}
  \bibinfo{person}{David~R. Cheriton}.} \bibinfo{year}{1989}\natexlab{}.
\newblock \showarticletitle{Leases: An Efficient Fault-Tolerant Mechanism for
  Distributed File Cache Consistency}. In \bibinfo{booktitle}{\emph{Proceedings
  of the Twelfth ACM Symposium on Operating Systems Principles}}.
  \bibinfo{publisher}{Association for Computing Machinery},
  \bibinfo{address}{New York, NY, USA}, \bibinfo{pages}{202--210}.
\newblock
\href{https://doi.org/10.1145/74850.74870}{doi:\nolinkurl{10.1145/74850.74870}}


\bibitem[Hagstr{\"o}m et~al\mbox{.}(2001)]%
        {hagstrom2001}
\bibfield{author}{\bibinfo{person}{{\AA}sa Hagstr{\"o}m},
  \bibinfo{person}{Sushil Jajodia}, \bibinfo{person}{Francesco
  Parisi-Presicce}, {and} \bibinfo{person}{Duminda Wijesekera}.}
  \bibinfo{year}{2001}\natexlab{}.
\newblock \showarticletitle{Revocations---A Classification}. In
  \bibinfo{booktitle}{\emph{Proceedings of the 14th IEEE Computer Security
  Foundations Workshop}}. \bibinfo{publisher}{IEEE Computer Society},
  \bibinfo{address}{Washington, DC, USA}, \bibinfo{pages}{44--58}.
\newblock
\href{https://doi.org/10.1109/CSFW.2001.930135}{doi:\nolinkurl{10.1109/CSFW.2001.930135}}


\bibitem[Khan(2026a)]%
        {stopmeansstop2026}
\bibfield{author}{\bibinfo{person}{Sajjad Khan}.}
  \bibinfo{year}{2026}\natexlab{a}.
\newblock \bibinfo{title}{Stop Means Stop: Measuring and Repairing the
  Enforcement Gap in Agent-Framework Control Primitives}.
\newblock
\showeprint[arxiv]{2607.14166}~[cs.SE]
\urldef\tempurl%
\url{https://arxiv.org/abs/2607.14166}
\showURL{%
\tempurl}
\newblock
\shownote{Version 3, revised 8 August 2026}.


\bibitem[Khan(2026b)]%
        {tokenbudgets2026}
\bibfield{author}{\bibinfo{person}{Sajjad Khan}.}
  \bibinfo{year}{2026}\natexlab{b}.
\newblock \bibinfo{title}{Token Budgets: An Empirical Catalog of 63 {LLM}-Agent
  Budget-Overrun Incidents, with an Affine-Typed {Rust} Mitigation as a Case
  Study}.
\newblock
\showeprint[arxiv]{2606.04056}~[cs.SE]
\urldef\tempurl%
\url{https://arxiv.org/abs/2606.04056}
\showURL{%
\tempurl}
\newblock
\shownote{Version 1, submitted 2 June 2026}.


\bibitem[Li et~al\mbox{.}(2026)]%
        {crosssubstrate2026}
\bibfield{author}{\bibinfo{person}{Yang Li}, \bibinfo{person}{Sergey Volkov},
  \bibinfo{person}{Hai Liu}, \bibinfo{person}{Zongsi Xu}, \bibinfo{person}{Xiyu
  Chen}, \bibinfo{person}{Tuo Zhou}, \bibinfo{person}{Dian Shao},
  \bibinfo{person}{Hao Sun}, {and} \bibinfo{person}{Ye Lu}.}
  \bibinfo{year}{2026}\natexlab{}.
\newblock \bibinfo{title}{Beyond Agent Harnesses: Cross-Substrate Authority for
  Multi-Agent Systems}.
\newblock
\showeprint[arxiv]{2609.08472}~[cs.MA]
\urldef\tempurl%
\url{https://arxiv.org/abs/2609.08472}
\showURL{%
\tempurl}
\newblock
\shownote{Version 1, submitted 8 September 2026}.


\bibitem[Lin et~al\mbox{.}(2026)]%
        {selfevolvingsafety2026}
\bibfield{author}{\bibinfo{person}{Ruixiao Lin}, \bibinfo{person}{Xinhao Deng},
  \bibinfo{person}{Qingming Li}, \bibinfo{person}{Jianan Ma},
  \bibinfo{person}{Yunhao Feng}, \bibinfo{person}{Yuqi Qing},
  \bibinfo{person}{Zhenyuan Li}, \bibinfo{person}{Yechao Zhang},
  \bibinfo{person}{Shiwen Cui}, \bibinfo{person}{Changhua Meng},
  \bibinfo{person}{Tianwei Zhang}, \bibinfo{person}{Xingjun Ma},
  \bibinfo{person}{Qi Li}, \bibinfo{person}{Ke Xu}, {and}
  \bibinfo{person}{Shouling Ji}.} \bibinfo{year}{2026}\natexlab{}.
\newblock \bibinfo{title}{Safety in Self-Evolving {LLM} Agent Systems: Threats,
  Amplification, and Case Studies}.
\newblock
\showeprint[arxiv]{2606.23075}~[cs.CR]
\urldef\tempurl%
\url{https://arxiv.org/abs/2606.23075}
\showURL{%
\tempurl}
\newblock
\shownote{Version 1, submitted 22 June 2026}.


\bibitem[Liu and Yu(2026)]%
        {attestingdelegation2026}
\bibfield{author}{\bibinfo{person}{Lifei Liu} {and} \bibinfo{person}{Haoran
  Yu}.} \bibinfo{year}{2026}\natexlab{}.
\newblock \bibinfo{title}{Attesting Outputs and Delegation Ancestry in
  Multi-Agent {AI} Systems}.
\newblock
\showeprint[arxiv]{2608.30387}~[cs.CR]
\urldef\tempurl%
\url{https://arxiv.org/abs/2608.30387}
\showURL{%
\tempurl}
\newblock
\shownote{Version 1, submitted 31 August 2026}.


\bibitem[Liu et~al\mbox{.}(2026)]%
        {vera2026}
\bibfield{author}{\bibinfo{person}{Lifei Liu}, \bibinfo{person}{Haoran Yu},
  {and} \bibinfo{person}{Xiaochong Jiang}.} \bibinfo{year}{2026}\natexlab{}.
\newblock \bibinfo{title}{{VERA}: Authority-Preserving Edge Revocation for
  Federated {AI}-Agent Workflows}.
\newblock
\showeprint[arxiv]{2608.30091}~[cs.AI]
\urldef\tempurl%
\url{https://arxiv.org/abs/2608.30091}
\showURL{%
\tempurl}
\newblock
\shownote{Version 1, submitted 30 August 2026}.


\bibitem[Mazzocchetti(2026)]%
        {runtimegovernance2026}
\bibfield{author}{\bibinfo{person}{Adam Mazzocchetti}.}
  \bibinfo{year}{2026}\natexlab{}.
\newblock \bibinfo{title}{Runtime Governance for Agentic {AI}: Action-Boundary
  Control with Trusted Provenance and Fail-Closed Execution}.
\newblock
\showeprint[arxiv]{2608.16891}~[cs.AI]
\urldef\tempurl%
\url{https://arxiv.org/abs/2608.16891}
\showURL{%
\tempurl}
\newblock
\shownote{Version 1, submitted 17 May 2026}.


\bibitem[Muruaga(2026)]%
        {boundedagents2026}
\bibfield{author}{\bibinfo{person}{Xabier Muruaga}.}
  \bibinfo{year}{2026}\natexlab{}.
\newblock \bibinfo{title}{Bounded Agents: Delegation Security for Multi-Agent
  {AI} Systems}.
\newblock
\showeprint[arxiv]{2608.15888}~[cs.AI]
\urldef\tempurl%
\url{https://arxiv.org/abs/2608.15888}
\showURL{%
\tempurl}
\newblock
\shownote{Version 1, submitted 16 August 2026}.


\bibitem[Saltzer and Schroeder(1975)]%
        {saltzer1975}
\bibfield{author}{\bibinfo{person}{Jerome~H. Saltzer} {and}
  \bibinfo{person}{Michael~D. Schroeder}.} \bibinfo{year}{1975}\natexlab{}.
\newblock \showarticletitle{The Protection of Information in Computer Systems}.
\newblock \bibinfo{journal}{\emph{Proc. IEEE}} \bibinfo{volume}{63},
  \bibinfo{number}{9} (\bibinfo{year}{1975}), \bibinfo{pages}{1278--1308}.
\newblock
\href{https://doi.org/10.1109/PROC.1975.9939}{doi:\nolinkurl{10.1109/PROC.1975.9939}}


\bibitem[Samuel et~al\mbox{.}(2010)]%
        {tuf2010}
\bibfield{author}{\bibinfo{person}{Justin Samuel}, \bibinfo{person}{Nick
  Mathewson}, \bibinfo{person}{Justin Cappos}, {and} \bibinfo{person}{Roger
  Dingledine}.} \bibinfo{year}{2010}\natexlab{}.
\newblock \showarticletitle{Survivable Key Compromise in Software Update
  Systems}. In \bibinfo{booktitle}{\emph{Proceedings of the 17th ACM Conference
  on Computer and Communications Security}}. \bibinfo{publisher}{Association
  for Computing Machinery}, \bibinfo{address}{New York, NY, USA},
  \bibinfo{pages}{61--72}.
\newblock
\href{https://doi.org/10.1145/1866307.1866315}{doi:\nolinkurl{10.1145/1866307.1866315}}


\bibitem[Santos-Grueiro(2026)]%
        {effectclosure2026}
\bibfield{author}{\bibinfo{person}{Igor Santos-Grueiro}.}
  \bibinfo{year}{2026}\natexlab{}.
\newblock \bibinfo{title}{When Does Authorization End? Effect Closure at
  Provider Boundaries}.
\newblock
\showeprint[arxiv]{2609.02866}~[cs.CR]
\urldef\tempurl%
\url{https://arxiv.org/abs/2609.02866}
\showURL{%
\tempurl}
\newblock
\shownote{Version 1, submitted 2 September 2026}.


\bibitem[Sengupta(2026)]%
        {sea2026}
\bibfield{author}{\bibinfo{person}{Biswa Sengupta}.}
  \bibinfo{year}{2026}\natexlab{}.
\newblock \bibinfo{title}{Self-Evolving Agents with Anytime-Valid
  Certificates}.
\newblock
\showeprint[arxiv]{2607.00871}~[cs.AI]
\urldef\tempurl%
\url{https://arxiv.org/abs/2607.00871}
\showURL{%
\tempurl}
\newblock
\shownote{Version 1, submitted 1 July 2026}.


\bibitem[South et~al\mbox{.}(2025)]%
        {authenticateddelegation2025}
\bibfield{author}{\bibinfo{person}{Tobin South}, \bibinfo{person}{Samuele
  Marro}, \bibinfo{person}{Thomas Hardjono}, \bibinfo{person}{Robert Mahari},
  \bibinfo{person}{Cedric~Deslandes Whitney}, \bibinfo{person}{Dazza
  Greenwood}, \bibinfo{person}{Alan Chan}, {and} \bibinfo{person}{Alex
  Pentland}.} \bibinfo{year}{2025}\natexlab{}.
\newblock \bibinfo{title}{Authenticated Delegation and Authorized {AI} Agents}.
\newblock
\showeprint[arxiv]{2501.09674}~[cs.CY]
\urldef\tempurl%
\url{https://arxiv.org/abs/2501.09674}
\showURL{%
\tempurl}
\newblock
\shownote{Version 1, submitted 16 January 2025}.


\bibitem[{\textcommabelow S}tef{\u a}nescu and Aciob{\u a}ni{\textcommabelow
  t}ei(2026)]%
        {lineageuma2026}
\bibfield{author}{\bibinfo{person}{Stefania {\textcommabelow S}tef{\u a}nescu}
  {and} \bibinfo{person}{Iulian Aciob{\u a}ni{\textcommabelow t}ei}.}
  \bibinfo{year}{2026}\natexlab{}.
\newblock \showarticletitle{Lineage: A {UMA} 2.0 Protocol Extension for
  Multi-Hop {LLM} Agent Delegation}.
\newblock \bibinfo{journal}{\emph{IEEE Access}}  \bibinfo{volume}{14}
  (\bibinfo{year}{2026}), \bibinfo{pages}{104574--104591}.
\newblock
\href{https://doi.org/10.1109/ACCESS.2026.3708618}{doi:\nolinkurl{10.1109/ACCESS.2026.3708618}}


\bibitem[Tallam(2026)]%
        {authprop2026}
\bibfield{author}{\bibinfo{person}{Krti Tallam}.}
  \bibinfo{year}{2026}\natexlab{}.
\newblock \bibinfo{title}{Authorization Propagation in Multi-Agent {AI}
  Systems: Identity Governance as Infrastructure}.
\newblock
\showeprint[arxiv]{2605.05440}~[cs.AI]
\urldef\tempurl%
\url{https://arxiv.org/abs/2605.05440}
\showURL{%
\tempurl}
\newblock
\shownote{Version 1, submitted 6 May 2026}.


\bibitem[Talluri et~al\mbox{.}(2026)]%
        {topologybudget2026}
\bibfield{author}{\bibinfo{person}{Abhijit Talluri}, \bibinfo{person}{Pujith
  Anne}, \bibinfo{person}{Bhagavan~Choudary Pendiyala}, {and}
  \bibinfo{person}{Raghavendra Chilukuri}.} \bibinfo{year}{2026}\natexlab{}.
\newblock \bibinfo{title}{Retrieval-Conditioned Topology Selection with
  Provable Budget Conservation for Multi-Agent Code Generation}.
\newblock
\showeprint[arxiv]{2605.05657}~[cs.AI]
\urldef\tempurl%
\url{https://arxiv.org/abs/2605.05657}
\showURL{%
\tempurl}
\newblock
\shownote{Version 1, submitted 7 May 2026}.


\bibitem[Tong et~al\mbox{.}(2026)]%
        {aidguard2026}
\bibfield{author}{\bibinfo{person}{Yingzhe Tong}, \bibinfo{person}{Leyu Dai},
  {and} \bibinfo{person}{Songhui Guo}.} \bibinfo{year}{2026}\natexlab{}.
\newblock \bibinfo{title}{{AID-Guard}: Stateful Authorization for Delegated
  Agent Effects}.
\newblock
\showeprint[arxiv]{2608.21159}~[cs.CR]
\urldef\tempurl%
\url{https://arxiv.org/abs/2608.21159}
\showURL{%
\tempurl}
\newblock
\shownote{Version 1, submitted 21 August 2026}.


\bibitem[Torres-Arias et~al\mbox{.}(2019)]%
        {intoto2019}
\bibfield{author}{\bibinfo{person}{Santiago Torres-Arias},
  \bibinfo{person}{Hammad Afzali}, \bibinfo{person}{Trishank~Karthik
  Kuppusamy}, \bibinfo{person}{Reza Curtmola}, {and} \bibinfo{person}{Justin
  Cappos}.} \bibinfo{year}{2019}\natexlab{}.
\newblock \showarticletitle{in-toto: Providing Farm-to-Table Guarantees for
  Bits and Bytes}. In \bibinfo{booktitle}{\emph{28th USENIX Security
  Symposium}}. \bibinfo{publisher}{USENIX Association}, \bibinfo{address}{Santa
  Clara, CA}, \bibinfo{pages}{1393--1410}.
\newblock
\showISBNx{978-1-939133-06-9}
\urldef\tempurl%
\url{https://www.usenix.org/conference/usenixsecurity19/presentation/torres-arias}
\showURL{%
\tempurl}


\bibitem[Wang et~al\mbox{.}(2026a)]%
        {regenharness2026}
\bibfield{author}{\bibinfo{person}{Kailin Wang}, \bibinfo{person}{Haoxiang
  Jie}, \bibinfo{person}{Yaoyuan Yan}, \bibinfo{person}{Zhiyou Heng}, {and}
  \bibinfo{person}{Zhaosong Li}.} \bibinfo{year}{2026}\natexlab{a}.
\newblock \bibinfo{title}{{RegenHarness}: A Robot Agent Harness with
  Evidence-Gated Recursive Self-Improvement}.
\newblock
\showeprint[arxiv]{2609.27612}~[cs.RO]
\href{https://doi.org/10.48550/arXiv.2609.27612}{doi:\nolinkurl{10.48550/arXiv.2609.27612}}
\newblock
\shownote{Version 1, submitted 23 September 2026}.


\bibitem[Wang(2026)]%
        {prv2026}
\bibfield{author}{\bibinfo{person}{Molly Wang}.}
  \bibinfo{year}{2026}\natexlab{}.
\newblock \bibinfo{title}{Spawn Freely, Act Sparingly: Progressive Risk Vesting
  for Recursive {LLM}-Agent Trees}.
\newblock
\showeprint[arxiv]{2609.01035}~[cs.AI]
\urldef\tempurl%
\url{https://arxiv.org/abs/2609.01035}
\showURL{%
\tempurl}
\newblock
\shownote{Version 1, submitted 1 September 2026}.


\bibitem[Wang et~al\mbox{.}(2026b)]%
        {harnesstampering2026}
\bibfield{author}{\bibinfo{person}{Xing Wang}, \bibinfo{person}{Xiaoyi Zhang},
  {and} \bibinfo{person}{Jie Shao}.} \bibinfo{year}{2026}\natexlab{b}.
\newblock \bibinfo{title}{Auditing Harness Tampering in Self-Improving Agents}.
\newblock
\showeprint[arxiv]{2609.00069}~[cs.CL]
\urldef\tempurl%
\url{https://arxiv.org/abs/2609.00069}
\showURL{%
\tempurl}
\newblock
\shownote{Version 1, submitted 30 August 2026}.


\bibitem[Ye and Tan(2026)]%
        {agentcontracts2026}
\bibfield{author}{\bibinfo{person}{Qing Ye} {and} \bibinfo{person}{Jing Tan}.}
  \bibinfo{year}{2026}\natexlab{}.
\newblock \bibinfo{title}{Agent Contracts: A Formal Framework for
  Resource-Bounded Autonomous {AI} Systems}.
\newblock
\showeprint[arxiv]{2601.08815}~[cs.MA]
\urldef\tempurl%
\url{https://arxiv.org/abs/2601.08815}
\showURL{%
\tempurl}
\newblock
\shownote{Version 3, revised 25 March 2026}.


\bibitem[Zhang et~al\mbox{.}(2025)]%
        {dgm2025}
\bibfield{author}{\bibinfo{person}{Jenny Zhang}, \bibinfo{person}{Shengran Hu},
  \bibinfo{person}{Cong Lu}, \bibinfo{person}{Robert Lange}, {and}
  \bibinfo{person}{Jeff Clune}.} \bibinfo{year}{2025}\natexlab{}.
\newblock \bibinfo{title}{Darwin {G\"odel} Machine: Open-Ended Evolution of
  Self-Improving Agents}.
\newblock
\showeprint[arxiv]{2505.22954}~[cs.AI]
\urldef\tempurl%
\url{https://arxiv.org/abs/2505.22954}
\showURL{%
\tempurl}
\newblock
\shownote{Version 3, revised 12 March 2026}.


\bibitem[Zhang et~al\mbox{.}(2026)]%
        {autogenesis2026}
\bibfield{author}{\bibinfo{person}{Wentao Zhang}, \bibinfo{person}{Zhe Zhao},
  \bibinfo{person}{Haibin Wen}, \bibinfo{person}{Yingcheng Wu},
  \bibinfo{person}{Cankun Guo}, \bibinfo{person}{Ming Yin}, {and}
  \bibinfo{person}{Bo An}.} \bibinfo{year}{2026}\natexlab{}.
\newblock \bibinfo{title}{Autogenesis: A Self-Evolving Agent Protocol}.
\newblock
\showeprint[arxiv]{2604.15034}~[cs.AI]
\urldef\tempurl%
\url{https://arxiv.org/abs/2604.15034}
\showURL{%
\tempurl}
\newblock
\shownote{Version 5, revised 20 June 2026}.


\bibitem[Zhang(2026)]%
        {agentlibos2026}
\bibfield{author}{\bibinfo{person}{Yingqi Zhang}.}
  \bibinfo{year}{2026}\natexlab{}.
\newblock \bibinfo{title}{Agent {libOS}: A Runtime Substrate for
  Capability-Controlled Self-Evolving {LLM} Agents}.
\newblock
\showeprint[arxiv]{2606.03895}~[cs.OS]
\urldef\tempurl%
\url{https://arxiv.org/abs/2606.03895}
\showURL{%
\tempurl}
\newblock
\shownote{Version 3, revised 18 August 2026}.


\bibitem[Zhang and Zhang(2026)]%
        {earnedauthority2026}
\bibfield{author}{\bibinfo{person}{Zhaoxi Zhang} {and} \bibinfo{person}{Xiaomei
  Zhang}.} \bibinfo{year}{2026}\natexlab{}.
\newblock \bibinfo{title}{Are You Still the Agent I Authorized? Earned
  Authority under a Fixed Ceiling for Evolving Agents}.
\newblock
\showeprint[arxiv]{2607.23586}~[cs.AI]
\urldef\tempurl%
\url{https://arxiv.org/abs/2607.23586}
\showURL{%
\tempurl}
\newblock
\shownote{Version 1, submitted 26 July 2026}.


\bibitem[Zhao and Zhao(2026)]%
        {ourark2026}
\bibfield{author}{\bibinfo{person}{Roy Zhao} {and} \bibinfo{person}{Zhenyu
  Zhao}.} \bibinfo{year}{2026}\natexlab{}.
\newblock \bibinfo{title}{Code Is the Body: Agent-Owned Software Bodies for
  Recursive Evolution and Descent}.
\newblock
\showeprint[arxiv]{2607.28691}~[cs.SE]
\urldef\tempurl%
\url{https://arxiv.org/abs/2607.28691}
\showURL{%
\tempurl}
\newblock
\shownote{Version 1, submitted 30 July 2026}.


\bibitem[Zheng et~al\mbox{.}(2026)]%
        {acrfence2026}
\bibfield{author}{\bibinfo{person}{Yusheng Zheng}, \bibinfo{person}{Yiwei
  Yang}, \bibinfo{person}{Wei Zhang}, {and} \bibinfo{person}{Andi Quinn}.}
  \bibinfo{year}{2026}\natexlab{}.
\newblock \bibinfo{title}{{ACRFence}: Preventing Semantic Rollback Attacks in
  Agent Checkpoint-Restore}.
\newblock
\showeprint[arxiv]{2603.20625}~[cs.CR]
\urldef\tempurl%
\url{https://arxiv.org/abs/2603.20625}
\showURL{%
\tempurl}
\newblock
\shownote{Version 1; CoDAIM Workshop 2026}.


\bibitem[Zhou(2026)]%
        {dynamiccapabilities2026}
\bibfield{author}{\bibinfo{person}{Ziling Zhou}.}
  \bibinfo{year}{2026}\natexlab{}.
\newblock \bibinfo{title}{Governing Dynamic Capabilities: Cryptographic Binding
  and Reproducibility Verification for {AI} Agent Tool Use}.
\newblock
\showeprint[arxiv]{2603.14332}~[cs.CR]
\urldef\tempurl%
\url{https://arxiv.org/abs/2603.14332}
\showURL{%
\tempurl}
\newblock
\shownote{Version 2, revised 19 March 2026}.


\bibitem[Zhu et~al\mbox{.}(2026)]%
        {badgenius2026}
\bibfield{author}{\bibinfo{person}{Guojun Zhu}, \bibinfo{person}{Xunheng
  Huang}, \bibinfo{person}{Peng Yin}, \bibinfo{person}{Jiahui Xie},
  \bibinfo{person}{Sanguo Zhang}, {and} \bibinfo{person}{Doudou Zhou}.}
  \bibinfo{year}{2026}\natexlab{}.
\newblock \bibinfo{title}{Bad Genius: Counterfactual-Guided Harness Evolution
  Beyond Task-Specific Shortcuts}.
\newblock
\showeprint[arxiv]{2609.18366}~[cs.AI]
\href{https://doi.org/10.48550/arXiv.2609.18366}{doi:\nolinkurl{10.48550/arXiv.2609.18366}}
\newblock
\shownote{Version 3, revised 24 September 2026}.


\bibitem[Zhu and Wang(2026)]%
        {zhu2026quiescence}
\bibfield{author}{\bibinfo{person}{Genliang Zhu} {and} \bibinfo{person}{Chu
  Wang}.} \bibinfo{year}{2026}\natexlab{}.
\newblock \bibinfo{title}{Authorization Revocation for Long-Running {AI}
  Agents: Root-Scoped Quiescence under Delegation and Asynchronous Execution}.
\newblock
\showeprint[arxiv]{2609.21284}~[cs.PL]
\href{https://doi.org/10.48550/arXiv.2609.21284}{doi:\nolinkurl{10.48550/arXiv.2609.21284}}
\newblock
\shownote{Version 1, submitted 18 September 2026}.


\end{thebibliography}

\end{document}